\documentclass[11pt]{article}

\def\ds{\displaystyle}

\usepackage{amsmath}
\usepackage{amsfonts}
\usepackage{amssymb}\usepackage{amsmath}
\usepackage{amsfonts}
\usepackage{amssymb}

\begin{document}
\pagestyle{headings}
\renewcommand{\thefootnote}{\alph{footnote}}

\title{Introduction to quantum field theory exhibiting interaction}
\author{Glenn Eric Johnson\\Oak Hill, VA.\\E-mail: glenn.e.johnson@gmail.com}
\maketitle

{\bf Abstract:} This note is an introduction to methods of construction for Hilbert space realizations of relativistic quantum physics. The realizations satisfy a revision to Wightman's functional analytic axioms and exhibit interaction in physical spacetimes. The local commutativity, relativistic invariance, positive energy and Hilbert space realization axioms are satisfied. The revision eliminates conjecture that a real quantum field is necessarily a Hermitian Hilbert space operator. The resulting explicit scattering cross sections coincide with the first contributing order from Feynman series for a neutral scalar field.


\section{Introduction}
Extrapolation of the successful methods of ordinary quantum mechanics to relativistic quantum physics achieves phenomenological success in the Feynman series although there is no demonstration that the developments are consistent with quantum mechanics [\ref{pct},\ref{bogo},\ref{wightman-hilbert}]. Lacking is a demonstration that the states of a quantum field theory (QFT) of interest are realized as elements of a Hilbert space. It has not been feasible to either display Hilbert space realizations of QFT that exhibit interaction in physical spacetimes nor to demonstrate that such QFT can not be realized. Efforts to better characterize relativistic quantum physics include description of quantum fields in the language of functions [\ref{pct},\ref{bogo},\ref{wight},\ref{borchers}]. The original Wightman-functional development of QFT [\ref{wight},\ref{borchers}] is that the properties of a quantum field are determined by a continuous linear functional dual to sequences of Schwartz test functions [\ref{gel2}]. The sequences of Schwartz functions are an involutive algebra. The involution provides that a multiplication in the algebra is realized as a Hermitian Hilbert space operator that is recognized as the quantum field. The flaw of the Wightman-functional development is that only physically trivial realizations have been demonstrated. However, either weakening the local commutativity condition [\ref{baum},\ref{lechner1},\ref{lechner2}] or alternatives to the Schwartz functions [\ref{gej05}] admit realizations of relativistic quantum physics with interaction in physical spacetimes. While both variations lead to realizations, there are additional compelling reasons to question the selection of the Schwartz functions. Selection of the Schwartz functions implies Hermitian Hilbert space field operators but an observable field does not necessarily correspond to a Hermitian Hilbert space field operator.

This note discusses constructions of Wightman-functionals on alternative sets of functions. The admission of alternative sets of functions is the only revision; the Wightman axioms for local commutativity, relativistic invariance, positive energy and the Hilbert space realization of quantum mechanics are satisfied. The constructions demonstrate that the conjecture that real quantum fields are Hermitian Hilbert space operators precludes realizations of relativistic quantum physics. Except for Hermitian Hilbert space field operators, the constructions display the established characteristics of relativistic quantum physics. Realization of quantum fields as Hermitian Hilbert space operators is unsettled although it is an established convention. Fields are observable but not all observable quantities are Hermitian Hilbert space operators. The original Wightman axioms exclude this possibility for quantum fields. The assertion that fields are Hermitian Hilbert space operators is despite counterexamples to the necessity of a correspondence of observable quantities with Hermitian Hilbert space operators, and despite the general deficiencies in the correspondence of real classical dynamic quantities with Hermitian operators [\ref{vonN}]. Counterexamples include symmetric products of nonrelativistic self-adjoint operators, for example, the formally Hermitian operator corresponding to ${\bf x}^3{\bf p}$ [\ref{bogo},\ref{johnson}], and the incompatibility of relativity with a Hermitian operator for ${\bf x}$ [\ref{johnson},\ref{yngvason-lqp},\ref{wigner}]. Selection of the Schwartz functions constrains the functional analytic development of QFT to result in unbounded, Hermitian Hilbert space field operators [\ref{pct}]. Here, this constraint is considered an unnecessary limitation on a functional analytic development. Indeed, constructions of random processes associated with analytic extensions of Wightman-functionals [\ref{agw},\ref{iqf}] suggest that the persistent lack of Hilbert space realizations for relativistic quantum physics of interest results from this constraint, that the lack of realizations is due to incompatibility of local, relativistic interaction with Hermitian Hilbert space field operators. Free fields provide no guidance in this regard. The constructions satisfy a solvable variant of the functional analysis problem posed by Wightman.

The necessity of Hermitian Hilbert space field operators is eliminated by basing the Hilbert space construction upon a non-involutive algebra of function sequences. The constructed Wightman-functionals provide a semi-norm for function sequences that are specialized to solve the functional analysis problem and are unconstrained by conjecture that quantum fields are Hermitian Hilbert space operators. The Wightman-functional is dual to an algebra designated as ${\cal A}$ and the component functions have Fourier transforms that are Schwartz tempered test functions of the momenta and infinitely differentiable functions of the energies. Slow growth with the energies is permitted and as a consequence, ${\cal A}$ includes generalized functions of time as well as the spacetime Schwartz functions. For the constructions, Lorentz invariance effectively reduces the dimensionality of the Wightman-functionals. Study of the constructions reduces to consideration of generalized functions in momentum coordinates dual to Schwartz functions of one less dimension than spacetime. ${\cal A}$ includes the functions used by Lehmann, Symanzik and Zimmermann in the calculation of scattering amplitudes [\ref{bogo}]. The Hilbert space is constructed from a subalgebra ${\cal B}\subset {\cal A}$ that labels the positive energy states. In the case of a free field, a semi-norm on positive energy states extends beyond ${\cal B}$ to an involutive algebra and results in Hermitian Hilbert space free field operators but the necessity of an extension is eliminated in the revision.

In this study, the constructions are for neutral, Lorentz scalar fields that exhibit particles in interaction. The development here is limited to a single Lorentz scalar field to simplify this introductory development and isolate unconventional properties of the constructions from the considerations of additional realizations of the Lorentz group. More general cases are constructed in [\ref{gej05}]. In this note, the constructions are designated as UQFT, unconstrained QFT, to distinguish revised axioms and the constructions from the original Wightman or Haag-Kastler (algebraic) QFT developments. The constructions consider the generalized functions\begin{equation}\label{gen-form} T(p_1,p_2,\ldots p_n)=\delta(p_1+p_2+\ldots p_n)\, \delta(p_1^2\!-\!m^2)\,\delta(p_2^2\!-\!m^2)\ldots \delta(p_n^2\!-\!m^2)\end{equation}suggested by the random process approach to construction of Wightman-functionals. These forms implement Poincar\'{e} covariance, the mass shell singularities that imply interaction, and a semi-norm that provides the Hilbert space realization. Forms based on (\ref{gen-form}) define continuous linear functionals in three or more spacetime dimensions when $m>0$, and four or more dimensions permits $m=0$. Consideration of a range of random process constructions [\ref{ieee}] results in no evident alternatives to generalized functions based on the form (\ref{gen-form}) for an explicit functional analytic development of local, relativistic quantum physics. But, (\ref{gen-form}) is excluded by the original Wightman axioms. Symmetric forms implement local commutativity, or limitations on the energy support implement the spectral support condition, but both local commutativity and positive energy are not achievable with spacetime Schwartz functions. Satisfaction of the axioms for relativistic quantum physics is achieved by a construction of physical states from the subalgebra ${\cal B}$ of functions with zeros on negative energy mass shells. This revision to the Wightman-functional development follows consideration that there are no physical states of negative energy and as a consequence, there is no necessity for labels of physical states with negative energy. The functions in ${\cal B}$ suffice to label the physical states. In analogy with the observation that Wightman-functionals are generalized functions in more than four dimensions (generalized functions of four spacetime arguments also define generalized functions for more than four arguments), the revised Wightman-functionals are defined for all values of the energies but satisfy the physically motivated Wightman conditions for positive energies. Like functions in four dimensions, functions selected to contribute only on positive energies suffice to label the physical states. The physical states are the elements of the constructed Hilbert space.

The approximation of cross sections [\ref{feymns}] suggests that Feynman series results are asymptotic to results from quantum mechanics. For weak coupling, the UQFT constructions approximate cross sections derived by the Feynman rules. These constructions result in non-forward scattering amplitudes\[\langle (p_1,\ldots p_n)^{\mathit{in}}|(p_{n+1},\ldots p_{n+m})^{\mathit{out}} \rangle = c_{n+m}\;\delta(p_1 \ldots\!+\!p_n \!-\!p_{n+1}\ldots \!-\!p_{n+m})\]that are proportional with a coefficient $i$ to the first contributing order of a Feynman series [\ref{weinberg}] with an interaction Hamiltonian density of $H_{int}(x)=\sum a_k :\!\Phi(x)^k\!:\;$ with $k\geq 4$ and $a_k=c_k\,(2\pi)^{2k-4}/k!$ with the $c_k$ equal to moments of a nonnegative measure.

The constructions consist of an expanded algebra of function sequences ${\cal A}$, a specialized subalgebra ${\cal B}\subset {\cal A}$ that labels the elements of a Hilbert space realization of relativistic quantum physics, and Wightman-functionals dual to ${\cal A}$ that provide the semi-norm for ${\cal B}$. After a development of notation, the functions in ${\cal A}$ and ${\cal B}$ are defined in sections \ref{sec-A} and \ref{sec-B}. Revised axioms A1-A5 for UQFT are described in section \ref{sec-ax} using Borchers' scalar QFT development. Section \ref{sec-suffcnt} develops sufficient conditions for a Wightman-functional to satisfy the revised axioms A2-A5. The explicit realizations of the revised axioms are developed in section \ref{realiz} and satisfaction of A1 is demonstrated. The study concludes with evaluation of the scattering amplitudes, and demonstrations that the vacuum is in a one-dimensional subspace of translational-invariant states and ${\cal B}$ includes no functions of bounded spatial support. The physically trivial free fields, archetypes for relativistic quantum physics, are included in both QFT and UQFT. Free fields are singularly removed from UQFT exhibiting interaction in a sense discussed in section \ref{conclu}. Additional comments on the physical and technical implications of the revised axioms are included in section \ref{conclusion}.

Variations of the constructions include: charges; multiple particle species; additional representations of the Lorentz group (higher spins, bosons and fermions) [\ref{gej05}]; massless particles [\ref{mp01}]; and approximations to Feynman series cross sections for Compton scattering in electrodynamics [\ref{feymns}].

\section{Revised axioms}
In this section, Borchers' development of scalar quantum fields that satisfy the Wightman axioms [\ref{borchers}] is used to describe the functional analytic development of relativistic quantum physics and discuss the revisions that distinguish UQFT. Both the Wightman and UQFT axioms describe a continuous linear functional, the Wightman-functional, that is local and Poincar\'{e} covariant, and provides a semi-norm that results in a Hilbert space realization of states with positive energy. The distinction between QFT and UQFT is in the selection of algebras of function sequences.

\subsection{General definitions}\label{sec-defn}
Notation covers the description of functions and generalized functions with multiple, four dimensional spacetime arguments. Spacetime coordinates are designated $x:=t,{\bf x}$ and energy-momentum vectors are $p:=E,{\bf p}$. $x,p \in {\bf R}^4$, ${\bf x},{\bf p}\in {\bf R}^3$. $x,p$ are Lorentz vectors. $x^2:=t^2-{\bf x}^2$, $p^2:=E^2-{\bf p}^2$ and $px:=Et-{\bf p}\!\cdot\!{\bf x}$ use the Minkowski signature and ${\bf x}^2$ is the square of the Euclidean length $\|{\bf x}\|^2={\bf x}\cdot{\bf x}$ in ${\bf R}^3$. Multiple arguments include an identification index. Ascending or descending sequences of multiple arguments are denoted $(x)_{j,k}:= x_j,x_{j+1},\ldots x_k$ and $(x)_{j,k}:= x_j,x_{j-1},\ldots x_k$ respectively. $(x)_n:=(x)_{1,n}$.\[\omega_j^2:=\omega({\bf p}_j)^2:=m^2+{\bf p}_j^2\]with a mass $m>0$ and $E_j^2=\omega_j^2$ describe mass shells in ${\bf R}^4$. $E_j=\omega_j$ is the positive mass shell and $E_j=-\omega_j$ is the negative mass shell. $\tilde{f}((p)_n)$ denotes the Fourier transform of $f((x)_n)$. The Fourier transform adopted here is the evident multiple spacetime argument extension of\[\tilde{f}(p):= \int \frac{dx}{(2\pi)^2}\; e^{-ipx} f(x)\]and $\tilde{T}(\tilde{f}):=T(f)$. Summation notation is used for generalized functions,\[\int dx\; T(x) f(x):=T(f)\]for a generalized function $T(x)$ and a function $f(x)\in {\cal A}$ with $x\in {\bf R}^4$ in this single argument case. In particular, the Dirac delta is $\int ds\; \delta(s-t)f(s):=f(t)$ and the first derivative is $\int ds\; \dot{\delta}(s-t)f(s):=-\dot{f}(t)$ using $\dot{f}(t)$ to designate the derivative and with $s,t \in {\bf R}$.

The Hilbert space operator terms used are Hermitian, symmetric, and self-adjoint: an operator $A$ with domain ${\cal D}_A$ in a Hilbert space with scalar product $\langle u|v \rangle$ is {\em Hermitian} if $\langle u|Av\rangle= \langle Au|v \rangle$ for every $u,v \in {\cal D}_A$. Hermiticity is necessary to {\em symmetry} (dense ${\cal D}_A$) and {\em self-adjointness} (dense ${\cal D}_A={\cal D}_{A^*}$).

\subsection{The algebra ${\cal A}$}\label{sec-A}
\newcounter{defintn}
\setcounter{defintn}{1}
\renewcommand{\thedefintn}{D.\arabic{defintn}}
{\em Definition} \thedefintn: The algebra ${\cal A}$ consists of terminating sequences of functions $f_n((x)_n)$. The sequences are denoted\[\underline{f}:=(f_0,f_1(x_1),f_2(x_1,x_2)\ldots,f_n((x)_n),\ldots).\]The Fourier transforms of the component functions $f_n((x)_n)$ are Schwartz tempered test functions of the momenta ${\bf p}_j$ when the energies $E_j$ are evaluated on mass shells, $E_j=\pm \omega({\bf p}_j)$. $f_0 \in {\bf C}$. Notation generally neglects to distinguish function sequences from their component functions, for example, $f_n((x)_n) \in {\cal A}$ is an abbreviated description for $f_n((x)_n)$ is the $n$ spacetime argument function from a sequence $\underline{f}\in {\cal A}$.

Addition in ${\cal A}$ is\[\underline{f}+\underline{g}=(f_0+g_0,\ldots, f_n((x)_n)+g_n((x)_n),\ldots)\]and the product is\begin{equation}\label{prod} \underline{f} \,{\bf x}\, \underline{g} := (f_0g_0, \ldots, \sum_{\ell=0}^n f_{\ell}((x)_{\ell})\,g_{n-\ell}((x)_{\ell+1,n}), \dots).\end{equation}Equipped with these two operations, ${\cal A}$ is a unital, associative algebra over the complex numbers [\ref{borchers}]. $\underline{1} \,{\bf x}\, \underline{f}=\underline{f} \,{\bf x}\, \underline{1}=\underline{f}$ for $\underline{1}=(1,0,0\ldots)$.

In the original development of Wightman QFT, the function sequences consist of Schwartz test functions, $f_n((x)_n)\in {\cal S}({\bf R}^{4n})$ [\ref{pct},\ref{borchers},\ref{gel2}], and the algebra of sequences of Schwartz test functions is designated ${\bf \Sigma}$. For UQFT, an expanded set of functions ${\cal A}$ is used. The functions in ${\cal A}$ include Schwartz tempered test functions, ${\bf \Sigma} \subset{\cal A}$, and as a consequence, ${\cal A}$ includes functions of bounded support in spacetime as well as functions with Fourier transforms of bounded energy-momentum support [\ref{gel2}].

Only particular generalized functions are considered for UQFT. The generalized functions of interest are generalized functions of the momenta and one of a finite number of particular generalized functions of the energies. The generalized function $T((x)_n)$ dual to functions $f_n((x)_n)\in{\cal A}$ are limited to\begin{equation}\label{emsupp}\renewcommand{\arraystretch}{1.25} \begin{array}{rl} T(f_n) &=\tilde{T}(\tilde{f}_n)\\
 &:= {\ds \sum_{(s)_n}} {\ds \int} d(p)_n\; {\ds \prod_{j=1}^b} \delta(E_j-s_j \omega_j)\, \tilde{T}_{(s)_n}(({\bf p})_n) \tilde{f}_n((p)_n)\\
 &={\ds \sum_{(s)_n}} {\ds \int} d({\bf p})_n\; \tilde{T}_{(s)_n}(({\bf p})_n) \tilde{f}_n((s \omega, {\bf p})_n)\end{array}\end{equation}with the summation over the $2^n$ possibilities for the signs $s_j=\pm 1$ of the $n$ energies and with a possibly distinct generalized function $\tilde{T}_{(s)_n}(({\bf p})_n)$ for each term. The generalized functions $\tilde{T}_{(s)_n}(({\bf p})_n)$ do not include derivatives with respect to the momenta. To define $T(f_n)$, it is sufficient that the Fourier transforms with $4n$ arguments $\tilde{f}_n((p)_n)$ are Schwartz functions of the momenta $({\bf p})_n \in{\bf R}^{3n}$ when the energies are on a mass shell, $E_j^2=\omega_j^2$.\begin{equation}\label{a-defn}f_n((x)_n) \in {\cal A} \qquad\mbox{if}\qquad \tilde{f}_n((s \omega,{\bf p})_n) \in S({\bf R}^{3n})\end{equation}with each $s_j=\pm1$. (\ref{a-defn}) suffices when the $\tilde{T}_{(s)_n}(({\bf p})_n)\in {\cal S}'({\bf R}^{3n})$, the generalized functions dual to the set of Schwartz tempered test functions ${\cal S}({\bf R}^{3n})$. The generalized functions $\tilde{T}((p)_n)$ are solutions to the Klein-Gordon equation that include the case of a real, scalar free field.

A construction for ${\cal A}$ results from noting that if $\tilde{g}((p)_n)$ is a multiplier in ${\cal S}'({\bf R}^{4n})$, then the $\tilde{g}((s \omega,{\bf p})_n)$ are multipliers in ${\cal S}'({\bf R}^{3n})$. The condition (\ref{a-defn}) is satisfied in the space formed as the span of the infinitely differentiable functions\begin{equation}\label{a-struct}\tilde{f}_n((p)_n):=\tilde{g}((p)_n) \tilde{\varphi}(({\bf p})_n)\end{equation}with $\tilde{\varphi}(({\bf p})_n) \in S({\bf R}^{3n})$ and $\tilde{g}((p)_n)$ a multiplier in ${\cal S}'({\bf R}^{4n})$ [\ref{gel2}]. The topology of ${\cal A}$ is determined by the countable norms for spatial Schwartz functions ${\cal S}({\bf R}^{3n})$ and the finite number of possible signs $s_j=\pm1$. (\ref{a-struct}) satisfies (\ref{a-defn}) as a consequence of the linear growth and infinite differentiability of $\omega_j$ in the momenta ${\bf p}_j$ when $m>0$. If for all $p\in {\bf R}^4$, a polynomially bounded growth $C_N:=C_N({\bf p})$ and an integer $N$,\[ |\tilde{g}(p)|<C_N\,(1+E^2)^N \qquad \mbox{then}\qquad |\tilde{g}(\pm \omega,{\bf p})| <C_N\,(1\!+\!m^2\!+\! \|{\bf p}\|^2)^N\]for all ${\bf p}\in {\bf R}^3$. $\tilde{g}(\pm \omega,{\bf p})$ is continuously infinitely differentiable as a result of the infinite differentiability of the $\omega_j$ and the chain rule, and then the polynomial growth bound provides that $\tilde{g}(\pm \omega,{\bf p})$ is a multiplier in ${\cal S}'({\bf R}^{3})$ [\ref{gel2}].

The functions of spacetime resulting from (\ref{a-struct}) are inverse Fourier transforms as generalized functions, convolutions\[f_n((x)_n)=\int \frac{d({\bf y})_n}{(2\pi)^{\frac{3n}{2}}}\; g((t,{\bf y})_n)\varphi(({\bf x}-{\bf y})_n) \in {\cal A},\]of the inverse Fourier transform as a generalized function of the multiplier $\tilde{g}((p)_n) \in {\cal S}'({\bf R}^{4n})$ and the inverse Fourier transform of a tempered test function $\tilde{\varphi}(({\bf p})_n)\in {\cal S}({\bf R}^{3n})$. The $f_n((x)_n)$ are tempered test functions when $\tilde{g}((p)_n) \in {\cal S}({\bf R}^{4n})\subset {\cal S}'({\bf R}^{4n})$. For the LSZ functions introduced in section \ref{sec-scat},\[\renewcommand{\arraystretch}{1.75} \begin{array}{rl} \ell_1(x_1;0)&:={\ds \int \frac{dp}{(2\pi)^2}}\; e^{ipx_1} (\omega+E) \tilde{\varphi}({\bf p})\\
 &= \delta(t_1){\ds \int \frac{d{\bf p}}{2\pi}}\;e^{-i{\bf p}\cdot {\bf x}_1} \,\omega\, \tilde{\varphi}({\bf p}) -i\dot{\delta}(t_1){\ds \int \frac{d{\bf p}}{2\pi}}\;e^{-i{\bf p}\cdot {\bf x}_1} \tilde{\varphi}({\bf p}),\end{array}\]are Schwartz functions of ${\bf x}_1$ and generalized functions with point support in time. From (\ref{a-struct}), $\tilde{\ell}_1(p;0)\in {\cal A}$ when $\tilde{\varphi}({\bf p})\in {\cal S}({\bf R}^3)$.
\newline

\addtocounter{defintn}{+1}
{\em Definition} \thedefintn: Complex conjugation with argument order reversal defines an automorphism of ${\cal A}$. The $*$-map $\underline{f}\mapsto \underline{f}^*$ is defined by the mapping of component functions\[f_n((x)_n) \mapsto f_n^*((x)_n) := \overline{f_n}((x)_{n,1}).\]$\overline{f_n}((x)_n)$ indicates the complex conjugate of the complex-valued function that has real arguments $(x)_n$. This $*$-map is an automorphism of ${\cal A}$ as a consequence of that neither real nor imaginary function components nor particular spacetime arguments are distinguished in the definition of ${\cal A}$. The $*$-map is an involution of ${\cal A}$ since it is an automorphism and satisfies $\underline{f}^{**}=\underline{f}$, $(\underline{g}+\underline{f})^* =\underline{g}^* + \underline{f}^*$ and $(\underline{g} \,{\bf x}\, \underline{f})^* =\underline{f}^* \,{\bf x}\, \underline{g}^*$. The Fourier transform of the $*$-mapped function is related to the Fourier transform of $f_n((x)_n)$ by\begin{equation}\label{mapf}\widetilde{f^*}_n((p)_n) = \overline{\tilde{f}_n} (-p_n, -p_{n-1}, \ldots ,-p_1).\end{equation}$\widetilde{f^*}_n((p)_n)$ designates the Fourier transform of $f_n^*((x)_n)$, distinct from the $*$-mapping of $\tilde{f}_n((p)_n)$ and the notation is unambiguous with the convention that the $*$-map is considered only for functions on spacetime.
\newline

\addtocounter{defintn}{+1}
{\em Definition} \thedefintn: An automorphism of ${\cal A}$ implements Poincar\'{e} transformations.\begin{equation}\label{poin-trans}(a,\Lambda)\underline{f}:=(f_o, \ldots f_n(\Lambda^{-1}(x_1-a), \ldots \Lambda^{-1}(x_n-a)),\ldots )\end{equation}with $\Lambda$ a proper orthochronous Lorentz transformation of a Lorentz vector and $a$ a constant Lorentz vector. The automorphism for the Fourier transformed functions is\begin{equation}\label{poin-p-trans}(a,\Lambda)\underline{\tilde{f}}:=(f_o, \ldots \exp(i(p_1+p_2\ldots +p_n)a) \tilde{f}_n(\Lambda^{-1}p_1, \ldots \Lambda^{-1}p_n),\ldots ).\end{equation}

\subsection{The subalgebra ${\cal B}$}\label{sec-B}
\addtocounter{defintn}{+1}
{\em Definition} \thedefintn: ${\cal B}$ is a subalgebra of function sequences with component functions derived from the component functions of ${\cal A}$. The Fourier transform of an $f_n((x)_n) \in {\cal B}$ is defined\begin{equation}\label{B-defn}\tilde{f}_n((p)_n):=\prod_{k=1}^n (E_k+\omega_k)\,\tilde{\varphi}_n((p)_n)\end{equation}for every $\varphi_n((x)_n)\in {\cal A}$. $f_0\in {\bf C}$ and $\underline{1}\in {\cal B}$.

The slow growth, infinitely differentiable $\omega_k$ and $E_k$ are multipliers in ${\cal A}$ and as a consequence, every $\tilde{f}_n((p)_n)$ from (\ref{B-defn}) is an element of ${\cal A}$. Sums and products (\ref{prod}) preserve (\ref{B-defn}) and as a consequence ${\cal B}$ is an algebra. The desired property of ${\cal B}$ is that the Fourier transforms have zeros on the negative energy mass shells.\begin{equation}\label{b-supprt}\tilde{f}_n((E,{\bf p})_n)=0\quad \mbox{when}\quad E_k=-\omega_k\end{equation}for any $k\in \{1,\ldots n\}$.

${\cal B}$ is a proper subset of ${\cal A}$. Many elements of ${\cal A}$ are not in ${\cal B}$ and in particular, many of the elements of the set of $*$-mapped functions ${\cal B}^*\subset {\cal A}$ are not  in common with ${\cal B}$. The $*$-map (\ref{mapf}) is not an automorphism of ${\cal B}$. The $\widetilde{f^*}_n((p)_n)=\overline{\tilde{f}_n}((-p)_{n,1})$ vanish when any $E_k=\omega_k$. Any $\tilde{f}_n((p)_n)\in {\cal B}\cap {\cal B}^*$ vanishes when $p_k^2=m^2$. As a consequence, the $\tilde{f}_n((p)_n)\in {\cal B}\cap {\cal B}^*$ are equivalent to zero for the generalized functions of interest (\ref{emsupp}) with support limited to the mass shells. The exception is $f_0$. The contributing sequences in ${\cal B}\cap {\cal B}^*$ are $(f_0,0,\ldots)$ with a real $f_0$. Discussed below in section \ref{sec-fields}, the equivalence of real functions of a single argument with zero precludes Hermitian Hilbert space field operators. Real functions of a single argument in ${\cal B}$ are necessarily Fourier transforms of functions in ${\cal B}^* \cap {\cal B}$ from $f^*(x_1)=\overline{f}(x_1)$ for the $*$-map (\ref{mapf}).

Every $f(x)$ of a single argument in ${\cal A}$ decomposes as $f=g+h^*$ with $g,h \in {\cal B}$.\[\tilde{g}(p)=\frac{\omega+E}{2\omega}\,\tilde{f}(p) \qquad\mbox{and}\qquad \widetilde{h^*}(p)=\overline{\tilde{h}}(-p)=\frac{\omega-E}{2\omega}\,\tilde{f}(p).\]

Whether proper orthochronous Poincar\'{e} transformation (\ref{poin-trans}) is an automorphism of ${\cal B}$ reduces to whether the zeros on the negative energy mass shells are preserved. Proper orthochronous Poincar\'{e} transformations of Lorentz vectors $p_j$ preserve the sign of the energy and the invariance of $p_j^2$ provides that zeros at $E_j=-\omega({\bf p}_j)$ map to $E'_j=-\omega({\bf p}'_j)$ with the Poincar\'{e} transformation $p_j\mapsto p_j'$. As a consequence, (\ref{poin-trans}) is an automorphism of ${\cal B}$.

\subsection{The revised axioms}\label{sec-ax}
To eliminate conditions that might preclude Hilbert space realization in the case of interaction, the assertion that the Wightman-functional satisfies the spectral support condition and provides a semi-norm for sequences of Schwartz tempered test functions ${\cal S}({\bf R}^{4n})$ is eliminated from the original Wightman axioms [\ref{pct},\ref{bogo},\ref{wight},\ref{borchers}]. In the case of a UQFT with a single Lorentz scalar field, the Wightman-functional $\underline{W}$ satisfies:

\begin{itemize} 
\item[A1.] {\em Description}: $\underline{W}$ is a continuous linear functional dual to an algebra ${\cal A}$ of function sequences. ${\cal A}$ includes functions of bounded spacetime support and a subalgebra ${\cal B}\subset {\cal A}$ includes functions with Fourier transforms of bounded energy-momentum support.
\item[A2.] {\em Relativistic invariance}: $\underline{W}$ is invariant with proper ortho\-chronous Poincar\'{e} transformations, $\underline{W}((a,\Lambda)\underline{f})=\underline{W}(\underline{f})$ for $\underline{f}\in {\cal A}$, and the Poincar\'{e} transformations are an automorphism of ${\cal B}$.
\item[A3.] {\em Spectral support}: When $\underline{f},\underline{g}\in {\cal B}$, $\underline{W}(\underline{f}^*\,{\bf x}\,\underline{g})$ includes only contributions from positive energies.
\item[A4.] {\em Locality}: $\underline{W}$ satisfies local commutativity for elements within ${\cal A}$ of bounded spacetime support. Local commutativity is that $\underline{W}$ is symmetric with transpositions of adjacent, space-like separated arguments.
\item[A5.] {\em Hilbert space realization}: $\underline{W}$ provides a semi-norm for ${\cal B}$. $\underline{W}(\underline{f}^*\,{\bf x}\,\underline{f})\geq 0$ for $\underline{f} \in {\cal B}$.\end{itemize}This statement of axioms follows the original Wightman axioms except for introduction of the algebra ${\cal A}$ and subalgebra ${\cal B}$ from sections \ref{sec-A} and \ref{sec-B}. If a substitution ${\cal B}={\cal A}\mapsto {\bf \Sigma}$ is made, then A1-A5 are the original Wightman axioms as expressed by Borchers. The revisions are the expansion of ${\bf \Sigma}$ to ${\cal A}$ that includes the LSZ functions and the limitation of the spectral support condition and the semi-norm to ${\cal B}$. The revised axioms eliminate conjecture and therefore escape the possibility that the lack of realizations of relativistic quantum physics of interest is due to the false assertion that fields, as classical dynamic quantities, must be Hermitian Hilbert space operators.

\subsection{Quantum fields and the Wightman-functional}\label{sec-fields}
\addtocounter{defintn}{+1}
{\em Definition} \thedefintn: The quantum field is multiplication in the algebra ${\cal A}$ of function sequences. From (\ref{prod}),\begin{equation} \Phi(f) \underline{g} := \underline{f} \,{\bf x}\, \underline{g}\label{fcompos}\end{equation}for $\underline{f}=(0,f(x_1),0,\ldots)$ and with $\underline{f},\underline{g} \in {\cal A}$. The properties of the quantum field are determined by the Wightman-functional.
\newline

\addtocounter{defintn}{+1}
{\em Definition} \thedefintn: The Wightman-functional is a sequence of generalized functions,\[\underline{W}:=(1,W_1(x_1),W_2(x_1,x_2),\ldots,W_n((x)_n),\ldots),\]dual to ${\cal A}$. The components of the Wightman-functional are denoted the {\em n-point generalized functions} $W_n((x)_n)$.

(\ref{fcompos}) associates the field with arguments of the $n$-point generalized functions.\begin{equation}\label{vev-eq}\langle \Omega| \Phi(f_a)\Phi(f_b)\ldots \Phi(f_z) \Omega\rangle := W_n(g)\end{equation}with\[g((x)_n) =f_a(x_1)f_b(x_2)\ldots f_z(x_n).\]Introducing a formal field $\Phi(x)$, the $n$-point generalized functions can be considered to result from a formally Hermitian field.\begin{equation}\label{sesquis2} \renewcommand{\arraystretch}{1.25} \begin{array}{rl} W_n((x)_n) &= \langle \Omega| \Phi(x_1)\ldots \Phi(x_n) \Omega\rangle\\
 &= \langle \Phi(x_k) \ldots \Phi(x_1)\Omega| \Phi(x_{k+1})\ldots \Phi(x_n) \Omega\rangle \end{array}\end{equation}for $1\leq k< n$. $\langle \Omega| \Phi(x_1)\ldots \Phi(x_n) \Omega\rangle
 = \langle \Phi(x_n) \ldots \Phi(x_1)\Omega| \Omega\rangle$ is real in this neutral scalar field case. Satisfaction of the original Wightman axioms provides that the $n$-point generalized functions are vacuum expectation values (VEV) of Hilbert space operators $\Phi(f)$. But when interaction is present, the multiplication (\ref{fcompos}) in ${\cal A}$ is not necessarily identified with a Hilbert space operator and, for the constructed UQFT, $\Phi(f)$ is necessarily not Hermitian in the Hilbert space of positive energy states. This distinction is now developed.

The sesquilinear function on ${\cal A}\times {\cal A}$,\begin{equation}\label{sesquis}\renewcommand{\arraystretch}{2.25} \begin{array}{rl}\underline{W}(\underline{f}^* \,{\bf x}\, \underline{g}) &:={\ds \sum_{n,m} \int} d(p)_{n+m}\;\tilde{W}_{n+m}((p)_{n+m}) \widetilde{f^*}_n((p)_n)\, \tilde{g}_m((p)_{n+1,n+m})\\
 &= \left\langle {\ds \sum_m \int} d(y)_m\;f_m((y)_m) \,\Phi_{\hat{1}}\ldots \Phi_{\hat{m}} \,\Omega|\; {\ds \sum_n \int} d(x)_n\;g_n((x)_n)\,\Phi_1\ldots \Phi_n\,\Omega\right\rangle \end{array}\end{equation}provides the scalar product of elements in a Hilbert space. $\Phi_k := \Phi(x_k)$ and $\Phi_{\hat{k}} := \Phi(y_k)$. A Hilbert space with elements labeled by sequences $\underline{f}\in {\cal B}$ results when $\underline{W}(\underline{f}^* \,{\bf x}\, \underline{f})$ is positive semidefinite for the function sequences in ${\cal B}$. The Hilbert space realization follows from quotient space and completion methods for a semi-norm on a linear vector space [\ref{pct},\ref{dieud}], in this case, ${\cal B}$. The elements of the constructed Hilbert space are labeled by equivalence classes of function sequences, equivalent in the semi-norm provided by the nonnegative sesquilinear function.\begin{equation}\label{norm} \| \underline{f} \|_{\cal B}:= \sqrt{\underline{W}(\underline{f}^* \, {\bf x}\, \underline{f})}.\end{equation}The Hilbert space representation of states is the result of a bijective map of equivalence classes of elements $\underline{f}\in {\cal B}$ for the semi-norm (\ref{norm}) to a dense set of elements in the Hilbert space. This map,\begin{equation}\label{isometry}\langle \underline{f}|\underline{g}\rangle= \underline{W}(\underline{f}^* \,{\bf x}\, \underline{g}),\end{equation}is an isometry.

The scalar product (\ref{isometry}), product of function sequences (\ref{prod}) and the $*$-map of sequences (\ref{mapf}) provide that should the field be defined as a Hilbert space operator that the operator adjoint field would be the field evaluated for the $*$-mapped function.\begin{equation}\label{adfield}\Phi(f)^*=\Phi(f^*).\end{equation}The definition of an adjoint operator, (\ref{mapf}), (\ref{fcompos}) and (\ref{isometry}) provide that\[\renewcommand{\arraystretch}{1.25} \begin{array}{rl} \langle \underline{h} |\Phi(f)\underline{g}\rangle &=\langle \Phi(f)^* \underline{h} |\underline{g}\rangle\\
 &= \underline{W}(\underline{h}^*\,{\bf x}\,\underline{f}\,{\bf x}\,\underline{g}\rangle\\
 &= \underline{W}((\underline{f}^*\,{\bf x}\,\underline{h})^*\,{\bf x}\,\underline{g}\rangle\\
 &=\langle \Phi(f^*)\underline{h} | \underline{g}\rangle.\end{array}\]As a consequence, the field is Hermitian, $\Phi(f)^*=\Phi(f)$, for real functions $f(x)=\overline{f}(x)$. From the discussion of section \ref{sec-B}, any real function of a single argument in ${\cal B}$ is in the equivalence class of $f(x)=0$ and the quantum fields (\ref{fcompos}) are not realized as Hermitian operators in the Hilbert space constructed from ${\cal B}$.

Preservation of equivalence classes for the norm (\ref{norm}) is necessary to definition of the quantum field as a Hilbert space operator. That is, if $\|\underline{g}+\underline{h}\|_{\cal B}=\|\underline{g}\|_{\cal B}$, then\[ \|\Phi(f)(\underline{g}+\underline{h})\|_{\cal B}=\|\underline{f}\,{\bf x}\,(\underline{g}+\underline{h})\|_{\cal B} \]must equal $\|\Phi(f)\underline{g}\|_{\cal B}$ for (\ref{isometry}) to define a Hilbert space operator. Multiplication preserves equivalence classes when the null space is a left ideal, when $\|\underline{f}\,{\bf x}\,\underline{h}\|_{\cal B}=0$ is implied by $\|\underline{h}\|_{\cal B}=0$. The Cauchy-Schwarz-Bunyakovsky (CSB) inequality provides that\[\renewcommand{\arraystretch}{1.25} \begin{array}{rl} \|\underline{f}\,{\bf x}\,\underline{h}\|_{\cal B}^2 &= \underline{W}((\underline{f}\,{\bf x}\,\underline{h})^*\,{\bf x}\, \underline{f}\,{\bf x}\,\underline{h})\\
 &\leq \|((\underline{f}\,{\bf x}\,\underline{h})^* \,{\bf x}\,\underline{f})^*\|_{\cal B}\;\|\underline{h}\|_{\cal B}\end{array}\]if $((\underline{f}\,{\bf x}\,\underline{h})^* \,{\bf x}\,\underline{f})^* \in{\cal B}$. In this event, the CSB inequality follows from A5 and the null space is a left ideal. In the original Wightman axioms, every $\underline{g}^*\in{\bf \Sigma}$ when $\underline{g}\in {\bf \Sigma}$ and then the field (\ref{fcompos}) defines a Hermitian Hilbert space operator from the validity of A5 for ${\bf \Sigma}$. ${\bf \Sigma}$ includes a dense set of real functions. In contrast, the development in section \ref{sec-B} for a UQFT provides that ${\cal B}\cap {\cal B}^*$ is trivial and when $\underline{f},\underline{h}\in {\cal B}$, then\[((\underline{f}\,{\bf x}\,\underline{h})^* \,{\bf x}\,\underline{f})^*= \underline{f}^*\,{\bf x}\,\underline{f} \,{\bf x}\,\underline{h}\notin{\cal B}.\]For UQFT, nonnegativity of the forms (\ref{sesquis}) does not necessarily apply to sequences $\underline{g} =\underline{f}^*\,{\bf x}\,\underline{f} \,{\bf x}\,\underline{h}\in {\cal A}$. When the semi-norm extends to include such sequences $\underline{g}$, then the CSB inequality is available to demonstrate that equivalence classes are preserved.

The field preserves equivalence classes and defines a Hilbert space operator when $\|\underline{h}\|_{\cal B}=0$ implies that $\|\underline{f}\,{\bf x}\, \underline{h}\|_{\cal B}=0$, for example, when $\underline{W}(\underline{g}^*\,{\bf x}\,\underline{g})$ is nonnegative for $\underline{g}\in {\cal A}$. For the constructions in section \ref{realiz} below, it is demonstrated that $\underline{W}(\underline{f}^*\,{\bf x}\,\underline{f})\geq 0$ for $\underline{f}\in {\cal B}$ but whether $\underline{W}(\underline{f}^*\,{\bf x}\,\underline{f})$ is definite for $\underline{f}\in {\cal A}$ is not characterized. Both cases occur among the constructions. For the free field Wightman-functional, $\underline{W}(\underline{f}^*\,{\bf x}\,\underline{f})\geq 0$ when $\underline{f}\in {\cal A}$, and for $W_2=0$ with higher order connected functions that are non-zero, there are $\underline{f}\in {\cal A}$ with $\underline{W}(\underline{f}^*\,{\bf x}\,\underline{f})<0$. Nevertheless, for the constructions, there are no $f(x)$ other than $f(x)= 0$ with $\Phi(f)\underline{g} = \Phi(f)^* \underline{g}\in {\cal B}$. Section \ref{conclu} includes additional discussion of differences between a conventional quantum field and the constructions.

\section{Sufficient conditions for a Wightman-functional}\label{sec-suffcnt}
Before the construction of Wightman-functionals, sufficient conditions for a Wightman-functional to satisfy the UQFT axioms A2-A5 for the sequences of functions ${\cal A}$ and ${\cal B}$ from sections \ref{sec-A} and \ref{sec-B} are developed. These results either are well known or are from [\ref{gej05}] and are included here for notation and discussion.
\newline

\newcounter{theorems}
\setcounter{theorems}{1}
\renewcommand{\thetheorems}{T.\arabic{theorems}}
{\em Theorem} \thetheorems: If the support of the Fourier transform of each $W_n((x)_n)$ is limited to $p_1+p_2\ldots +p_n=0$, then the Wightman-functional is invariant to translations.

The constructed $\underline{W}$ satisfy the translation invariance from A2 if\[\underline{W}(\underline{f}) = \underline{W}((a,1) \underline{f})\]for the automorphism (\ref{poin-trans}) of ${\cal A}$. The theorem follows from (\ref{poin-p-trans}) and $\underline{\tilde{W}}(\underline{\tilde{f}})=\underline{W}(\underline{f})$ since if the support of the $\tilde{W}_n((p)_n)$ is limited to $p_1+p_2\ldots+p_n=0$, then $(a,1)\underline{\tilde{f}}=\underline{\tilde{f}}$.
\newline

\addtocounter{theorems}{1}
{\em Theorem} \thetheorems: If each $W_n((x)_n)$ is expressed in proper orthochronous Lorentz invariants, then the Wightman-functional is Lorentz invariant.

The theorem follows from $\underline{\tilde{W}}(\underline{\tilde{f}})=\underline{W}(\underline{f})$, the realization of generalized functions as summations [\ref{gel2}] and the automorphism (\ref{poin-p-trans}) of ${\cal A}$ with $a=0$. Proper orthochronous Lorentz transformations preserve the signs of energies $E_j$ for time-like energy-momentum vectors $p_j$ ($p_j^2>0$). $p_j^2$ and $\sum_j p_j=0$ are Lorentz invariants. Then, $\tilde{W}_n((\Lambda p)_n)=\tilde{W}_n((p)_n)$ results in the theorem.

Poincar\'{e} invariance results from the composition of Lorentz transformations with translations.
\newline

\addtocounter{theorems}{1}
{\em Theorem} \thetheorems: If the supports of the $n$-point generalized functions are limited to mass shells and to $p_1+p_2+\ldots p_n=0$, then a Wightman-functional with $n$-point generalized functions of the form (\ref{emsupp}) satisfies the spectral support condition A3.

The spectral support condition A3 is that every finite contribution to $\underline{W}(\underline{u}^*\,{\bf x}\,\underline{v})$ when $\underline{u},\underline{v} \in {\cal B}$ is from a set of energy-momentum that lies within ${\cal E}_n^+$.\[{\cal E}_n^+ :=\{(p)_n \;|\; p_n \in \overline{V}^+,\;p_{n-1}+p_n \in \overline{V}^+, \ldots,\; p_2+\ldots p_n \in \overline{V}^+,\;p_1+p_2+\ldots p_n=0\}.\]The set $\overline{V}^+:=\{p\;|\; p^2 \geq 0,\mbox{ and }E\geq 0\}$ is the closed forward cone. A3 is satisfied if $\underline{W}({\cal M}_{sp})=0$ for ${\cal M}_{sp}$ the linear subspace of functions with Fourier transforms $\tilde{f}_n((p)_n)\in {\cal A}$ that vanish together with their derivatives when $(p)_n\in {\cal E}_n^+$ [\ref{borchers}].

From section \ref{sec-A} and (\ref{B-defn}), ${\cal A}$ and ${\cal B}$ include functions with Fourier transforms of bounded energy-momentum support to evaluate the support of $\tilde{W}_n((p)_n)$. The Fourier transforms of $v_j((x)_j)\in {\cal B}$ have zeros on the negative energy mass shells, and from (\ref{mapf}), the Fourier transforms of $u_j^*((x)_j)\in {\cal B}^*$ have zeros on the positive energy mass shells. Satisfaction of the theorem results from the observation that for the constructions, $\underline{f}=0$ is the only function sequence in ${\cal M}_{sp}$ of the form $\underline{f}=\underline{u}^*\,{\bf x}\, \underline{v}$ when $\underline{v},\underline{u}\in {\cal B}$. The form of the $n$-point generalized functions (\ref{emsupp}) does not include derivatives with respect to the energies or momenta and, as a consequence, the zeros of $\tilde{u}_k((p)_k),\tilde{v}_{n-k}((p)_{n-k})$ apply to the evaluation of $W_n(u_k^* v_{n-k})$. For $u_k((x)_k),v_{n-k}((x)_{n-k}) \in{\cal B}$ and when the supports of the components of the Wightman-functionals are limited to mass shells, the joint support of products of $\tilde{W}_n((p)_n)$, $\widetilde{u^*}_k((p)_k)$ and $\widetilde{v}_{n-k}((p)_{k+1,n})$ is necessarily contained within ${\cal E}_n^+$. The joint support has each $p_k^2=m^2$, $E_j<0$ for $j\leq k$, $E_j>0$ for $j>k$ and $p_1+p_2+\ldots p_n=0$. As a consequence, $(p)_n\in {\cal E}_n^+$ follows from closure of the cone $\overline{V}^+$ with convex addition, either directly or using the translation invariance (energy-momentum conservation) to find that\[\sum_{j=\ell}^n p_j = -\sum_{j=1}^{\ell-1} p_j \in \overline{V}^+.\]The negatives of Lorentz vectors with $E_j<0$ and $p_j^2>0$ are in $\overline{V}^+$.

Satisfaction of A3 follows consideration that the physical values $T(f)$ are the evaluations of generalized functions labeled by functions $f(x)$. There is no necessity for labels of physical states with negative energy because there are no physical states of negative energy. As a generalized function, the constructed $\tilde{T}(p)$ are defined on negative energies and this definition is used to check the local commutativity condition. Locality is formulated as a property of generalized functions defined using functions of bounded spacetime support. Local commutativity is established in ${\cal A}$ that includes functions of bounded spacetime support and ${\cal A}$ includes the subset ${\cal B}$ of functions that label the physical states. As a consequence, satisfaction of locality applies in ${\cal B}$.
\newline

\addtocounter{theorems}{1}
{\em Theorem} \thetheorems: If the $n$-point generalized functions are symmetric with transpositions of arguments, then the Wightman-functional satisfies the local commutativity condition A4.

From [\ref{borchers}], a Wightman-functional satisfies A4 if $\underline{W}(I_c)=0$ with $I_c$ the linear subspace of ${\cal A}$ with a base of functions $f(x_1,\ldots x_n)$ that decompose as the difference\[f((x)_n)=g(x_1\dots x_{i-1},x_i \ldots x_k, x_{k+1}\ldots x_n)-g(x_1\dots x_{i-1},x_{i'} \ldots x_{k'}, x_{k+1}\ldots x_n)\]and $g((x)_n)=0$ if $x_j-x_{\ell}$ is time-like for all $j,\ell \in \{i,i+1,\ldots k\}$, $j\neq \ell$. $\{i'\dots k'\}$ is any permutation of $\{i\ldots k\}$. $x$ is time-like if $x^2>0$.

Unconditional symmetry of the $W_n((x)_n)$ is an evident implementation of local commutativity and consistency with the unsymmetrical spectral support condition A3 is realized because of the non-involutive selection for ${\cal B}$. The unsymmetrical Pauli-Jordan function $\Delta(x-y)$ is symmetrical with transposition of $x,y$ when $(x-y)^2<0$ and this peculiar property is significant for satisfaction of A1-A5 with an extension of the free field semi-norm from ${\cal B}$ to ${\cal A}$. The properties of $\Delta(x-y)$ permit local, Hermitian Hilbert space free field operators that satisfy the spectral support condition A3. The constructions of UQFT do not rely on this property of $\Delta(x-y)$ for satisfaction of A4.

The implementation of causality as local commutativity follows from consideration of commuting Hermitian Hilbert space operators as quantum mechanical observables [\ref{bogo}]. Local commutativity is preserved in UQFT for the implications such as the statistics (\ref{b-ident1}) of the states, developed in section \ref{sec-B-result}, and the reality of products of a real field as well as for the implications for causality. The unconditional symmetry of real $W_n((x)_n)$ results in real values for products of a real quantum field, a property not exhibited even for conventional free quantum fields. Indeed, for real symmetric $W_n((x)_n)$,\[\renewcommand{\arraystretch}{1.25} \begin{array}{rl} 0 &=\underline{W}(\underline{u}^*\,{\bf x}\, [\Phi_x,\Phi_y]\, \underline{u})\\
 &=\underline{W}(\underline{u}^*\,{\bf x}\, \Phi_x\Phi_y\, \underline{u})-\underline{W}(\underline{u}^*\,{\bf x}\, \Phi_y\Phi_x\, \underline{u})\\
 &=\underline{W}(\underline{u}^*\,{\bf x}\, \Phi_x\Phi_y \,\underline{u})-\underline{W}((\Phi_x\Phi_y \,\underline{u})^*\,{\bf x}\,\underline{u})\\
 &=\underline{W}(\underline{u}^*\,{\bf x}\, \Phi_x\Phi_y \,\underline{u})-\overline{\underline{W}(\underline{u}^*\,{\bf x}\, \Phi_x\Phi_y\, \underline{u})}\\
 &=2i\; \Im \underline{W}(\underline{u}^*\,{\bf x}\, \Phi_x\Phi_y \,\underline{u})\end{array}\]from (\ref{mapf}), (\ref{sesquis}) and (\ref{adfield}), with $\Phi_x$ the field labeled by a real test function delta sequence in ${\cal A}$ supported near $x$, and $\Im z$ is the imaginary part of a complex number $z$. In contrast, the Pauli-Jordan function is complex-valued when $x$ and $y$ have a time-like separation. $\Phi_x$ is not a Hermitian operator in the constructed Hilbert space but is a quantity of interest for classical limits. The result is that expected values of the product of real fields are real.

The introduction of generator functionals for the $n$-point generalized functions simplifies the demonstration of a semi-norm for function sequences from ${\cal B}$.
\newline

\addtocounter{theorems}{1}
{\em Theorem} \thetheorems: If the Fourier transforms of the $n$-point generalized functions as duals to functions $\underline{f}=\underline{u}^*\,{\bf x}\,\underline{v}$ with $\underline{u},\underline{v}\in {\cal B}$ are generated by a power series\begin{equation}\label{t-gen}\tilde{W}_{n+m}((p)_{n+m}) = \prod_{j=1}^{n+m} \frac{\ds d\;}{\ds d\alpha_j}\; \exp({\cal K}_{n,m}((\alpha,p)_{n+m}))\end{equation}evaluated at $(\alpha)_{n+m}=0$, and if the generator functional is a nonnegatively weighted summation of factored terms,\begin{equation}\label{t-factor}{\cal K}_{n,m}((\alpha,p)_{n+m}) =\int d\mu_{\cal K}(\zeta)\;\overline{A_n}(\zeta,(\alpha,-p)_{n,1})A_m(\zeta,(\alpha,p)_{n+1,n+m}),\end{equation}then the Wightman-functional satisfies A5 and provides a semi-norm for ${\cal B}$.

The generator functional derives from a polynomial ${\cal K}_{n,m}((\alpha,p)_{n+m})$ in the $(\alpha)_{n+m}$ with generalized functions as coefficients. The generator is the power series in $(\alpha)_n$ that results from exponentiation of the polynomial. The power series is used as a convenient organization for the combinatorics in the expressions for $W_n((x)_n)$ and since only a finite number of derivatives are evaluated, no convergence of the power series is considered.

The demonstration of a semi-norm is an elaboration on Schur's product theorem for Hadamard products [\ref{horn}]. The $*$-map (\ref{mapf}), the sesquilinear function (\ref{sesquis}) and (\ref{t-gen}) result in\begin{equation} \label{structgen}\renewcommand{\arraystretch}{1.25} \begin{array}{l} \underline{W}(\underline{f}^*\,{\bf x}\, \underline{f}) ={\ds \sum_{n,m= 0}^{\infty}} W_{n+m}(f_n^* f_m)\\
 \qquad= \left( {\ds \sum_{n=0}^\infty \int} d(p)_n \; \overline{\tilde{f}_n}((-p)_{n,1}
) \right)\left({\ds \sum_{m=0}^\infty \int} d(p)_{n+1,n+m} \; \tilde{f}_m((p)_{n+1,n+m}) \right) \times \\
 \qquad\qquad\qquad \left({\ds \prod_{j=1}^{n+m}} \frac{\ds d\;}{\ds d\alpha_j}\right) \;\exp({\cal K}_{n,m}((\alpha,p)_{n+m}))\\
 \qquad= \left( {\ds \sum_{n=0}^\infty \int} d(p)_n \; \overline{\tilde{f}_n}((p)_n)\;{\ds \prod_{j=1}^n} \frac{\ds d\;}{\ds d\alpha_j}
 \right)\left({\ds \sum_{m=0}^\infty \int} d(q)_m \; \tilde{f}_m((q)_m)\;{\ds \prod_{j=1}^m} \frac{\ds d\;}{\ds d\beta_j} \right) \times \\
 \qquad\qquad\qquad \exp({\cal K}_{n,m}((\alpha,-p)_{n,1},(\beta,q)_m)) 
\end{array} \end{equation}when $(\alpha)_{n+m}=0$. In the third line, $q_k:=p_{n+k}$ and $\beta_k:=\alpha_{n+k}$ term by term, and the first $n$ summations of $(p)_n$ are relabeled as $(-p)_{n,1}$. For $1\leq j \leq n$, the $\alpha_j$ are relabeled as $\alpha_{n+1-j}$ and $\prod_j d/d\alpha_j$ and the condition $(\alpha)_n=0$ are invariant to this relabeling.

Substitution of the absolutely convergent series for the exponential function\[\exp({\cal K}_{n,m})=\lim_{N\rightarrow \infty} \left(1+\frac{{\cal K}_{n,m}}{N}\right)^N\]provides a convenient demonstration that $\underline{W}(\underline{f}^*\,{\bf x}\, \underline{f})\geq 0$. With one summation $d\mu_{\cal K}(\zeta_j)$ for each factor ${\cal K}_{n,m}$, substitution of (\ref{t-factor}) in (\ref{structgen}) provides that\[\renewcommand{\arraystretch}{2.25} \begin{array}{l} 
\underline{W}(\underline{f}^*\,{\bf x}\, \underline{f}) \approx \left( {\ds \sum_{n=0}^\infty \int} d(p)_n\, \overline{\tilde{f}_n}((p)_n){\ds \prod_{j=1}^n} \frac{\ds d\;}{\ds d\alpha_j}
 \right)\left({\ds \sum_{m=0}^\infty \int} d(q)_m\, \tilde{f}_m((q)_m){\ds \prod_{j=1}^m} \frac{\ds d\;}{\ds d\beta_j} \right) \times \\
 \qquad\qquad \left(1+{\ds \frac{1}{N}\int} d\mu_{\cal K}(\zeta)\; \overline{A_n}(\zeta,(\alpha,p)_n) A_m(\zeta,(\beta,q)_m)\right)^N\end{array}\]and reorganization results in\[\renewcommand{\arraystretch}{2.25} \begin{array}{l} 
\underline{W}(\underline{f}^*\,{\bf x}\, \underline{f}) \approx |f_0|^2+{\ds \int} d\mu_{\cal K}(\zeta_1)\; \left| {\ds \sum_{n=1}^\infty \int} d(p)_n \; \tilde{f}_n((p)_n)\;{\ds \prod_{j=1}^n} \frac{\ds d\;}{\ds d\alpha_j}A_n(\zeta_1,(\alpha,p)_n)\;\right|^2\\
 \qquad\quad +{\ds \frac{N\!-\!1}{2N}} {\ds \iint} d\mu_{\cal K}(\zeta_1)d\mu_{\cal K}(\zeta_2)\, \left| {\ds \sum_{n=1}^\infty \int} d(p)_n \; \tilde{f}_n((p)_n)\;{\ds \prod_{j=1}^n} \frac{\ds d\;}{\ds d\alpha_j}A_n(\zeta_1)A_n(\zeta_2)\, \right|^2+\ldots \end{array} \]with the abbreviated notation $A_n(\zeta_1):=A_n(\zeta_1,(\alpha,p)_n)$ in the final line. Each term in the summation is nonnegative for every $N$ when the weight $d\mu_{\cal K}(\zeta)\geq 0$. The summations over $n,m$ and the expansion for $\exp(x)$ have a finite number of terms. In the last line, the terms with a $W_0$ were segregated. $W_0=1$ and terms $W_n(f_0^* f_n)=\overline{f_0}\, W_n(f_n)=0$ for $n\geq 1$ from translation invariance (conservation of energy, $\sum_j E_j=0$) noting that every $E_j>0$ for energies that are on mass shells and when $f_n((x)_n)\in {\cal B}$. The substantive sign change and reordering of the first $n$ energy-momenta $p_j$ is from (\ref{t-factor}).
 
Then, (\ref{t-gen}) and (\ref{t-factor}) imply that $\underline{W}(\underline{f}^*\,{\bf x}\, \underline{f})\geq 0$ when $\underline{f}\in{\cal B}$.

\section{A Wightman-functional}\label{realiz}
\subsection{The Wightman-functional realization}\label{sec-constr}
The development now turns to construction of $n$-point generalized functions that are symmetric with interchange of arguments, that depend only on Lorentz invariants, with Fourier transforms of support limited to mass shells and $p_1+p_2\ldots p_n=0$, and that are generated by a form (\ref{t-gen}) that factors as in (\ref{t-factor}) for function sequences from ${\cal B}$. This construction is realized by (\ref{w-defn}).
\newline

\addtocounter{defintn}{+1}
{\em Definition} \thedefintn: The Fourier transforms of the $n$-point generalized functions $W_n((x)_n)$ are a finite summation of generalized functions denoted $\tilde{V}_{k,n-k}((p)_n)$.\begin{equation}\label{w-defn} \renewcommand{\arraystretch}{2.25} \begin{array}{rl} \tilde{W}_n((p)_n) &:={\ds \sum_\pi \sum_{k=0}^n}\; {\ds \frac{\Theta_{k,n}(E_{\pi_1},\ldots E_{\pi_n})}{k!\, (n-k)!}}\;\tilde{V}_{k,n-k}(p_{\pi_1},\ldots p_{\pi_n})\\
 &:={\ds \sum_{k=0}^n}\; {\ds \frac{n!}{k!\, (n-k)!}}\;\hat{\bf S}[\Theta_{k,n}((E)_n)\tilde{V}_{k,n-k}((p)_n)]\end{array}\end{equation}with\begin{equation}\label{v-defn}\tilde{V}_{k,n-k}((p)_n):=\left({\ds \prod_{j=1}^n} \frac{\ds d\;}{\ds d\alpha_j}\right) \; {\cal G}_o((\alpha,p)_n){\cal G}_{k,n-k}((\alpha,p)_n)\end{equation}evaluated at $(\alpha)_n=0$ for $n\geq 1$. $W_0:=1=\widetilde{W}_0$. ${\cal G}_o((\alpha,p)_n)$ is the generator functional for the free field generalized functions and ${\cal G}_{k,n-k}((\alpha,p)_n)$ is the generator functional for the contributions of higher order, more than two argument, connected generalized functions. The generators of the $\tilde{V}_{k,n-k}((p)_n)$ are formal power series in $(\alpha)_m$ with generalized function coefficients as discussed below Theorem T.5. The functions $\Theta_{k,n}((E)_n)$ and the finite summations over permutations of argument order $\{\pi_1,\pi_2,\ldots \pi_n\}$ are defined below.
\newline

\addtocounter{defintn}{+1}
{\em Definition} \thedefintn: The {\em energy ordering functions} are \begin{equation}\label{eorder}\Theta_{k,n}((E)_n):=\prod_{j_1=1}^k \theta(-E_{j_1}) \;\prod_{j_2=k+1}^n \theta(E_{j_2})\end{equation}and $\Theta_{k,n}((E)_n)=1$ when $-E_j>0$ for $j\leq k$ and $E_j>0$ for $k<j\leq n$. $\Theta_{k,n}((E)_n)=0$ otherwise. $\theta(x)$ is the Heaviside step function. The $\Theta_{k,n}((E)_n)$ are proper orthochronous Lorentz invariant functions of the energies when the $p_j^2=m^2$.\newline

\addtocounter{defintn}{+1}
{\em Definition} \thedefintn: A normalized symmetrization with permutations of arguments is defined\begin{equation}\label{permutation} \hat{\bf S}[T((x)_n)\ldots U((x)_n)]:= \frac{1}{n!} \sum_{\pi} T(x_{\pi_1}, \ldots x_{\pi_n})\ldots U(x_{\pi_1}, \ldots x_{\pi_n})\end{equation}with the summation over the $n!$ permutations $\{\pi_1,\pi_2,\dots \pi_n\}$ of $\{1,2,\ldots n\}$. For illustration,\[\hat{\bf S}[T((x)_2)]= \frac{1}{2}(T(x_1,x_2)+T(x_2,x_1)).\]This summation results in $\tilde{W}_n((p)_n)$ that are symmetric with transpositions of arguments.

The polynomials that generate the contributions of the free field are\begin{equation}\label{free-gen} \ln\left({\cal G}_o((\alpha,p)_n)\right) := {\ds \sum_{k=1}^n \sum_{j=k+1}^n} \;\tilde{\Delta}(p_k,p_j)\,\alpha_k\alpha_j\end{equation}and result in the free field $n$-point generalized functions,\begin{equation}\label{wodefn-free} \renewcommand{\arraystretch}{1.25} \begin{array}{l} {\ds \prod_{j=1}^n} \frac{\ds d\;}{\ds d\alpha_j}\, {\cal G}_o((\alpha,p)_\ell)= \left\{ \renewcommand{\arraystretch}{1.25} \begin{array}{ll} {\ds \sum_{\mathit{pairs}}} \tilde{\Delta}(p_{i_1},p_{i_2})\ldots \tilde{\Delta}(p_{i_{2\jmath \!-\!1}},p_{i_{2\jmath}}) & \quad n=2\jmath\\
 0 & \quad n=2\jmath\!+\!1 \end{array} \right. \end{array}\end{equation}when $(\alpha)_n=0$ and $\jmath$ is a positive integer. The indicated sum is over all $(2\jmath)!/(2^\jmath \jmath!)$ distinct pairings of the integers 1 through $n$ without regard to order. The indices of the two-point generalized functions are in ascending index order, $i_{j_1}<i_{j_2}$ when $j_1<j_2$. The neutral, Lorentz scalar free field of finite mass is described by a Pauli-Jordan function with a Fourier transform,\begin{equation}\label{twopoint}\renewcommand{\arraystretch}{1.25} \begin{array}{rl} \tilde{\Delta}(p_1,p_2) &:= \delta(p_1+p_2)\; \delta^+(p_2)\\
&= \delta({\bf p}_1+{\bf p}_2)\; \sqrt{2\omega_1}\, \delta^-(p_1)\, \sqrt{2\omega_2}\, \delta^+(p_2)\end{array}\end{equation}with\[\delta^\pm(p):= \theta(\pm E)\delta(p^2\!-\!m^2).\]This Pauli-Jordan function is supported only on mass shells. $\Delta(x_1,x_2)$ is connected [\ref{bogo}]. From (\ref{w-defn}), $W_2((x)_2)$ is a real function,\[\tilde{W}_2(p_1,p_2)=\tilde{\Delta}(p_1,p_2)+\tilde{\Delta}(p_2,p_1).\]As a dual to ${\cal B}$, the non-zero contribution is from $\tilde{W}_2(p_1,p_2)=\tilde{\Delta}(p_1,p_2)$.

The polynomials that generate the contributions of the $n$-point ($n\geq 3$) connected functions are\begin{equation}\label{genr2} \ln\left({\cal G}_{k,n-k}((\alpha,p)_n)\right) :={\ds \int d\sigma(\lambda) \int \frac{du}{(2\pi)^4}}\; {\ds \prod_{\ell =1}^n} (a_{k \ell}\!+\! \lambda \alpha_\ell e^{-ip_\ell u}\, \hat{\delta}(p_\ell))\end{equation}when $1< k < n-1$.\begin{equation}\label{hat-delta}\hat{\delta}(p):=\delta(p^2\!-\!m^2)=\delta^-(p)+\delta^+(p),\end{equation}$d\sigma(\lambda)$ is a nonnegative measure with finite moments,\begin{equation}\label{cn}c_n:=\int d\sigma(\lambda)\; \lambda^n\end{equation}and\[a_{k \ell}=\left\{ \begin{array}{ll} 0\qquad &\mbox{if }\ell=k-1,k,k+1,k+2\\ 1&\mbox{otherwise}\end{array}\right.\]eliminates a divergence from terms that are quadratic in the $(\alpha)_n$. All contributing terms in (\ref{genr2}) are quartic or higher degree in the $(\alpha)_n$.
\newline

\addtocounter{defintn}{+1}
{\em Definition} \thedefintn: The {\em conjoined functions} ${^n\tilde{C}}_{k,\eta}(p_{i_1},\ldots p_{i_\eta})$ are generalized functions that include $\eta$ indices $\{i_1,i_2,\ldots i_\eta\}\subseteq \{1,2,\ldots n\}$ in their description. The arguments are ordered $i_{j_1}<i_{j_2}$ when $j_1<j_2$.

From\[\int ds\; \exp(its)=2\pi \delta(t)\]as a generalized function,\begin{equation}\label{c-defn}\renewcommand{\arraystretch}{1.25}\begin{array}{rl} {^n\tilde{C}}_{k,\eta}(p_{i_1},\ldots p_{i_\eta})&:= \left({\ds \prod_{j=1}^\eta} {\ds \frac{d\quad}{d\alpha_{i_j}}}\right) \ln\left({\cal G}_{k,n-k}((\alpha,p)_n)\right)\\
 &=c_\eta \, \delta(p_{i_1}+\ldots p_{i_\eta})\,{\ds \prod_{j=1}^\eta}\; \hat{\delta}(p_{i_j})\end{array}\end{equation}when $\{k-1,k,k+1,k+2\} \subseteq \{i_1,i_2,\ldots i_\eta\} \subseteq \{1,2,\ldots n\}$ and with $\hat{\delta}(p)$ from (\ref{hat-delta}).\[{^n\tilde{C}}_{k,\eta}(p_{i_1},\ldots p_{i_\eta}):=0\]otherwise, in particular when $\eta,n=0,1,2,3$ or $k=0,1,n-1,n$. Demonstrated below in section \ref{sec-linf}, the conjoined functions are generalized functions dual to ${\cal A}$ when spacetime has three or more dimensions in the case of a finite mass $m$, and inclusion of massless particles requires four or more dimensions [\ref{mp01}]. All ${^n\tilde{C}}_{k,\eta}(p_{i_1},\ldots p_{i_\eta})=0$ and interaction vanishes when the $c_n=0$. It is demonstrated below in lemma T.12 that the conjoined functions are connected.

From (\ref{c-defn}) and for $k\neq 0,1,n-1,n$,\[{^nC}_{k,n}((x)_n)=c_n \int \frac{du}{(2\pi)^4} \; \prod_{j=1}^n\Delta_1(u-x_j)\]defined in terms of the Pauli-Jordan function\[\Delta_1(x):=\int \frac{dp}{(2\pi)^2}\; e^{ipx}\; \delta(p^2\!-\!m^2).\]When mollified by convolution with test functions, these connected functions do not exhibit rapid decline in the spatial difference variables ${\bf x}_{j+1}\!-\!{\bf x}_j$, in contradiction to an implication of the original Wightman axioms (theorem 10-4 [\ref{bogo}], [\ref{ruelle}]). The slow decline with large $|{\bf x}_{j+1}\!-\!{\bf x}_j|$ is not summable on unbounded intervals. The conjoined functions (\ref{c-defn}) are admitted by the revised axioms.

Satisfaction of axioms A2-A4 is evident from T.1-T.4 for the construction (\ref{w-defn}). (\ref{wodefn-free}), (\ref{twopoint}), (\ref{genr2}) and (\ref{c-defn}) substituted in (\ref{w-defn}) provide that the $\tilde{W}_n((p)_n)$ depend on $p_j^2$, the signs of the $E_j$ for time-like Lorentz vectors, and include factors of $\delta(p_{i_1}+\ldots p_{i_k})$. This form is invariant with proper orthochronous Lorentz transformations and the assumptions of T.2 are satisfied. Every $p_j$ appears once and only once in a delta function factor. As a consequence, the supports of the $\tilde{W}_n((p)_n)$ are limited to $p_1+p_2+\ldots p_n=0$ and the assumptions of T.1 are satisfied. The supports of the $\tilde{W}_n((p)_n)$ are limited to the mass shells $p_j^2=m^2$ and satisfy the assumptions of T.3. The $\tilde{W}_n((p)_n)$ are symmetric with transpositions of arguments to satisfy the assumptions of T.4.

Satisfaction of A5 and A1 are demonstrated in sections \ref{sec-B-result} and \ref{sec-linf}.
\newline

\addtocounter{defintn}{+1}
{\em Definition} \thedefintn: A generalized function $T_n((x)_n)$ is designated as {\em connected} if\[ {\bf P}[T_n](f_k^* \,(\rho a,1) f_{n-k}) \rightarrow 0\]for all $1\leq k<n$ and space-like Lorentz vector $a$ ($a^2<0$), as $|\rho|$ grows without bound and for any permutation ${\bf P}$ of the arguments of $T_n((x)_n)$. The sum of the distinct connected terms in $W_n((x)_n)$ is denoted the {\em $n$-point connected function}, ${^CW}((x)_n)$.

The definition of translation of $\underline{f}$ by $-\rho a$ is from (\ref{poin-trans}), and\[{\bf P}[T_n]((x)_n):=T_n(x_{\pi_1},\ldots x_{\pi_n})\]for one of the $n!$ distinct permutations $\{\pi_1,\pi_2,\dots \pi_n\}$ of $\{1,2,\ldots n\}$. This definition provides that a connected generalized function becomes negligible when any proper subset of arguments becomes greatly space-like separated from the remaining arguments. This definition of $n$-point connected functions applies for the construction (\ref{w-defn}) that represents the $n$-point generalized functions $W_n((x)_n)$ as a finite number of terms consisting of products of connected generalized functions with factors that have no arguments in common. In this case, $n$-point connected functions ${^CW}((x)_n)$ are identified by evaluation of ${\bf P}[W_n](f_k^* \,(\rho a,1) f_{n-k})$ as $|\rho|$ grows without bound for each of the finite number of permutations and $k$, and subtraction of the distinct results from $W_n((x)_n)$ results in ${^CW}((x)_n)$.

\subsection{The $n$-point generalized functions}\label{sec-linkr}
The $n$-point generalized functions $W_n((x)_n)$ are a finite sum of products of conjoined functions that have no arguments in common.

From (\ref{v-defn}), the coefficients in the product ${\cal G}_o((\alpha,p)_n)\,{\cal G}_{k,n-k}((\alpha,p)_n)$ are the generalized functions $\tilde{V}_{k,n-k}((p)_n)$. The $\tilde{V}_{k,n-k}((p)_n)$ expand in conjoined functions ${^n\tilde{C}}_{k,\eta}(p_{i_1},\ldots p_{i_\eta})$ using the link-cluster identity [\ref{bogo},\ref{ruelle-link}]. With\[{\cal K}_{k,n-k}((\alpha,p)_n):=\ln({\cal G}_o((\alpha,p)_n){\cal G}_{k,n-k}((\alpha,p)_n)),\]repeated differentiation results in\begin{equation}\label{link-cluster}\renewcommand{\arraystretch}{1.25} \begin{array}{l} \left( {\ds \prod_{j=1}^n} \frac{\ds d\;}{\ds d\alpha_j}\right) \exp({\cal K}_{k,n-k}((\alpha,p)_n)))\\
 \quad = \exp({\cal K}_{k,n-k}((\alpha,p)_n)))\; {\ds \sum_{\ell=1}^n \sum_{\{I_j\} \in \rho_{\ell,n}}\; \prod_{j=1}^\ell}\left( \frac{\ds d\quad}{\ds d\alpha_{i_{j_1}}}\ldots \frac{\ds d\quad\;}{\ds d\alpha_{i_{j\eta_j}}}\, {\cal K}_{k,n-k}((\alpha,p)_n)\right).\end{array}\end{equation}This result is related to Fa\`{a} di Bruno's formula and is verified by induction [\ref{combin}]. $\rho_{\ell,n}$ is the set of all partitions of the integers $\{1,2,\ldots n\}$ into $\ell$ nonempty and non-intersecting subsets $I_1, \ldots I_\ell$. The number of such partitions is the Stirling number of the second kind [\ref{nbs}]. There are $\eta_j$ numbers $(i_{j_1},\ldots i_{j\eta_j})$ within each subset $I_j$, ordered by magnitude $(i_{j_1} < i_{j_2} < \ldots )$. $\sum_{j=1}^\ell \eta_j = n$. Substituting (\ref{v-defn}) and (\ref{c-defn}), when $(\alpha)_n=0$ this results in the link-cluster expansion (cf.~eqn.~10.29 [\ref{bogo}]).\begin{equation}\label{faadibruno}\tilde{V}_{k,n-k}((p)_n) = {\ds \sum_{\ell=1}^n \sum_{\{I_j\} \in \rho_{\ell,n}}\; \prod_{j=1}^\ell}\; {^n\tilde{C}}_{k,\eta_j}(p_{i_{j_1}},\ldots p_{i_{j\eta_j}}).\end{equation}In this expansion, ${^nC}_{k,2}(x_{i_1},x_{i_2}):=\Delta(x_{i_1},x_{i_2})$ when $i_1,i_2 \in \{1,\ldots n\}$ is substituted for the ${^nC}_{k,2}(x_{i_1},x_{i_2})=0$ from (\ref{c-defn}). Contributing $V_{k,n-k}((x)_n)$ include\[\renewcommand{\arraystretch}{1.25} \begin{array}{rl} \tilde{V}_{0,2}((p)_2) &=\tilde{V}_{1,1}((p)_2)=\tilde{V}_{2,0}((p)_2)=(12)\\
\tilde{V}_{2,2}((p)_4) &=(1234)+(12)(34)+(13)(24)+(14)(23)\\
\tilde{V}_{2,3}((p)_5) &= \tilde{V}_{3,2}((p)_5) = (12345)\\
\tilde{V}_{2,4}((p)_6) &=(123456)+(1234)(56)+(12)(34)(56)+\ldots (16)(25)(34)\\
 \tilde{V}_{3,3}((p)_6) &=(123456)+(2345)(16)+(12)(34)(56)+\ldots (16)(25)(34)\\
 \tilde{V}_{4,2}((p)_6) &=(123456)+(3456)(12)+(12)(34)(56)+\ldots (16)(25)(34).\end{array}\]The abbreviated notation is\[(i_1 \ldots i_\eta) :={^n\tilde{C}}_{k,\eta}(p_{i_1},\ldots p_{i_\eta})\]defined by (\ref{c-defn}) except that ${^n\tilde{C}}_{k,2}(p_{i_1},p_{i_2})=\tilde{\Delta}(p_{i_1},p_{i_2})$. There are $(2\ell)!/(2^\ell \ell!)$ distinct pairings of $2\ell$ indices in the free field contribution, 3 pairings for $2\ell=4$ and 15 pairings for $2\ell=6$. Evaluation of (\ref{w-defn}) results from multiplication of the $\tilde{V}_{k,n-k}((p)_n)$ by $\Theta_{k,n}((E)_n)$, summation over $k$, and symmetrization over arguments. The resulting $\tilde{W}_n((p)_n)$ can be expressed\begin{equation}\label{w-result}\renewcommand{\arraystretch}{1.25} \begin{array}{rl} \tilde{W}_4((p)_4) &= {\ds \sum_{\mathit{part}}} ((1234)+(13)(24)+(14)(23))\; \Theta_{2,4}\\
\tilde{W}_5((p)_5) &= {\ds \sum_{\mathit{part}}} (12345)\;(\Theta_{2,5} + \Theta_{3,5})\\
\tilde{W}_6((p)_6) &= {\ds \sum_{\mathit{part}}} (123456)\;(\Theta_{2,6} +\Theta_{3,4} +\Theta_{4,6})+ \frac{1}{9}\{(2345)(16)+(1345)(26)\\
 &\qquad\qquad +(1245)(36)+(2346)(15)+(1346)(25)+(1246)(35)\\
 &\qquad\qquad +(2356)(14)+(1356)(24)+(1256)(34)\}\;\Theta_{3,6}\\
 &\qquad\qquad +\{(14)(25)(36)+(14)(26)(35)+(15)(24)(36)+(15)(26)(34)\\
 &\qquad\qquad +(16)(24)(35)+(16)(25)(34)\}\; \Theta_{3,6}.\end{array}\end{equation}The summation\[\sum_{\mathit{part}} f((p)_n)\, \Theta_{k,n}((E)_n)\]is the sum over all transpositions of the arguments $(p)_n$ that correspond to partitions of $n$ objects into two distinct subsets, $\{i_1,\ldots i_k\}$ and its set complement, without regard to order within subsets. $E_{i_1},\ldots E_{i_k}<0$ and $E_{i_{k+1}},\ldots E_{i_n}>0$. For example,\[\sum_{\mathit{part}} \hat{\delta}_1 \hat{\delta}_2 \hat{\delta}_3\; \Theta_{1,3}= \delta^-_1 \delta^+_2 \delta^+_3+\delta^+_1 \delta^-_2 \delta^+_3+\delta^+_1 \delta^+_2 \delta^-_3\]from (\ref{eorder}) and with $\hat{\delta}_k=\delta^+_k+\delta^-_k$ from (\ref{hat-delta}).
\newline

\addtocounter{theorems}{1}
{\em Theorem} \thetheorems: The connected terms of $W_n((x)_n)$ are\begin{equation}\label{connctd}{^C\tilde{W}}_n((p)_n):=\hat{\bf S}[{\ds \sum_k \frac{n!\,\Theta_{k,n}((E)_n)}{k!(n-k)!}} \; {^n\tilde{C}}_{k,n}((p)_n)]\end{equation}

This relation follows from (\ref{w-defn}) and the cluster decomposition of the $V_{k,n-k}$ in (\ref{faadibruno}), assuming for this argument and verified in lemma T.12 in section \ref{sec-vac} that the conjoined functions ${^nC}_{k,n}((x)_n)$ are connected. The identity\[\hat{\bf S}[{\ds \sum_{k=0}^n \frac{n!\,\Theta_{k,n}((E)_n)}{k!(n-k)!}}]=\prod_{j=1}^n \left(\theta(-E_j)+\theta(E_j)\right)=1\]and the description of the conjoined functions (\ref{c-defn}) result in equivalent forms for (\ref{connctd}), for example,\[{^C\tilde{W}}_n((p)_n)=c_n \, \delta(p_1+\ldots p_n) \prod_{j=1}^n \hat{\delta}(p_j) \;\left(1-\hat{\bf S}[\Theta_{0,n}+\Theta_{1,n}+\Theta_{n-1,n}+\Theta_{n,n}]\right)\]using the evident shorthand for $\Theta_{k,n}((E)_n)$.

Evident in (\ref{w-result}), when there is interaction these ${^CW}_n((x)_n)$ are distinct from truncated functions ${^TW}_n((x)_n)$ defined recursively from $W_n((x)_n)$ using the link-cluster expansion ($s$-logarithm). Truncated functions are connected when the $W_n((x)_n)$ satisfy the original Wightman axioms [\ref{haagasym},\ref{ahh}].
\newline

\addtocounter{theorems}{1}
{\em Theorem} \thetheorems: The $W_n((x)_n)$ defined by (\ref{w-defn}) coincide with free field $W_n((x)_n)$ as duals to ${\cal B}$ when the interaction vanishes, $c_n=0$ in (\ref{cn}).

From (\ref{genr2}) when the $c_n=0$, ${\cal G}_{k,n-k}((\alpha,p)_n)=1$ and ${\cal G}_o((\alpha,p)_n)$ is independent of $k$. (\ref{w-defn}), (\ref{v-defn}) and (\ref{wodefn-free}) result in\begin{equation}\label{free-con}\tilde{W}_n((p)_n) =\hat{\bf S}[{\ds \sum_{\mathit{pairs}}} \tilde{\Delta}(p_{i_1},p_{i_2})\ldots \tilde{\Delta}(p_{i_{2\ell\!-\!1}},p_{i_{2\ell}})\left({\ds \sum_{k=0}^n \frac{n!}{k!(n\!-\!k)!}}\, \Theta_{k,n}((E)_n)\right)]\end{equation}for $n=2\ell$ and $\tilde{W}_n=0$ for odd $n$. (\ref{twopoint}) provides that each $E_{i_{2j}}>0$ and $E_{i_{2j-1}}<0$ for $j=1,2,\ldots \ell$. As a consequence, the only terms of (\ref{free-con}) that contribute have $k=\ell$, $i_{2j-1}\leq \ell$ and $i_{2j}>\ell$ for $j=1,2,\dots \ell$. The summation over all distinct pairs that contribute to $W_{2\ell}(f_\ell^* g_\ell)$ when $\underline{f},\underline{g}\in {\cal B}$ results in a form that is symmetric with transpositions of arguments $i_{j_1},i_{j_2}$ when both $i_{j_1},i_{j_2}\leq \ell$ or $i_{j_1},i_{j_2}> \ell$. From the zeros of elements of ${\cal B}$ (\ref{b-supprt}) and due to this symmetry with argument transpositions, the sum over permutations (\ref{permutation}) results in accumulation of a count of $(\ell!)^2$ identical contributing terms and as a result\[\tilde{W}_{2\ell}((p)_{2\ell}) ={\ds \sum_{\mathit{pairs}}} \tilde{\Delta}(p_{i_1},p_{i_2})\ldots \tilde{\Delta}(p_{i_{2\ell\!-\!1}},p_{i_{2\ell}})\Theta_{\ell,2\ell}((E)_{2\ell}).\]This is identified as the Fourier transform of $W_n((x)_n)$ for a free field (\ref{wodefn-free}) applicable to $\underline{W}(\underline{f}^* \,{\bf x}\,\underline{g})$ with $\underline{f},\underline{g} \in{\cal B}$.
 
\subsection{The semi-norm}\label{sec-B-result}
An identity is useful to the demonstration that the construction (\ref{w-defn}) satisfies A5, that $\underline{W}$ provides the semi-norm (\ref{norm}) for ${\cal B}$.
\newline

\addtocounter{theorems}{1}
{\em Lemma} \thetheorems: As duals to ${\cal B}$, the $W_n((x)_n)$ coincide with the $V_{k,n-k}((x)_n)$ from (\ref{v-defn}) evaluated for symmetric functions. For $f_k((x)_k),f_{n-k}((x)_{n-k})\in {\cal B}$,\begin{equation}\label{b-ident1}W_n(f_k^*f_{n-k})=V_{k,n-k}\left(\hat{\bf S}[f_k^*]\,\hat{\bf S}[f_{n-k}]\right).\end{equation}

The identity (\ref{b-ident1}) follows from (\ref{w-defn}) and the zeros from (\ref{b-supprt}).\[\renewcommand{\arraystretch}{1.75} \begin{array}{rl} W_n(f_k^*f_{n-k})&= {\ds \sum_{\ell=0}^n} {\ds \int} d(p)_n\;n!\,{\ds \frac{\hat{\bf S}[\Theta_{\ell,n}((E)_n)\tilde{V}_{k,n-k}((p)_n)]}{\ell!\, (n-\ell)!}}\, \widetilde{f^*}_k((p)_k)\tilde{f}_{n-k}((p)_{k+1,n})\\
 &= {\ds \int} d(p)_n\;n!\,{\ds \frac{\hat{\bf S}[\Theta_{k,n}((E)_n)\tilde{V}_{k,n-k}((p)_n)]}{k!\, (n-k)!}}\, \widetilde{f^*}_k((p)_k)\tilde{f}_{n-k}((p)_{k+1,n})\\
 &={\ds \int} d(p)_n\;n!\,{\ds \frac{\Theta_{k,n}((E)_n)\tilde{V}_{k,n-k}((p)_n)}{k!\, (n-k)!}}\, \hat{\bf S}[\widetilde{f^*}_k((p)_k)\tilde{f}_{n-k}((p)_{k+1,n})]\\
 &= {\ds \int} d(p)_n\;\tilde{V}_{k,n-k}((p)_n)\, \hat{\bf S}[\widetilde{f^*}_k((p)_k)]\hat{\bf S}[\tilde{f}_{n-k}((p)_{k+1,n})].\end{array}\]The only contributing terms in the second line have $\ell=k$ since the product of $\Theta_{\ell,n}((E)_n)$, $\widetilde{f^*}_k((p)_k)$ and $\tilde{f}_{n-k}((p)_{k+1,n})$ vanishes otherwise due to the zeros of elements of ${\cal B}$ on negative mass shells. The third line results from relabeling the arguments in each term of the summation $\hat{\bf S}[]$ from (\ref{permutation}). The final line results from the vanishing of any permutation of arguments that transposes an argument of $\widetilde{f^*}_k((p)_k)$ with an argument of $\tilde{f}_{n-k}((p)_{k+1,n})$ due to the factor of $\Theta_{k,n}((E)_n)$ and the negative energy zeros of ${\cal B}$. The factor of $\Theta_{k,n}((E)_n)$ is redundant with the factors of $\hat{\bf S}[\widetilde{f^*}_k((p)_k)]$ and $\hat{\bf S}[\tilde{f}_{n-k}((p)_{k+1,n})]$ in the final line. (\ref{b-ident1}) displays Bose-Einstein symmetry of the states and emphasizes that the argument labels are not particle labels. The particles are indistinguishable.

The identity (\ref{b-ident1}) and the generators of $\tilde{V}_{k,n-k}((p)_n)$ (\ref{v-defn}) provide that the construction satisfies the assumptions of T.5 with functions $\hat{\bf S}[\tilde{f}_n((p)_n]\in {\cal B}$.\begin{equation} \renewcommand{\arraystretch}{1.25} \begin{array}{l} \underline{W}(\underline{f}^*\,{\bf x}\, \underline{f}) ={\ds \sum_{n,m\geq 0}^{\infty}} V_{n,m}(\hat{\bf S}[f_n^*] \hat{\bf S}[f_m])\\
 \qquad= \left( {\ds \sum_{n=0}^\infty \int} d(p)_n \; \hat{\bf S}[\overline{\tilde{f}_n}((-p)_{n,1})
] \right)\left({\ds \sum_{m=0}^\infty \int} d(p)_{n+1,n+m} \; \hat{\bf S}[\tilde{f}_m((p)_{n+1,n+m})] \right) \times \\
 \qquad\qquad\qquad \left({\ds \prod_{j=1}^{n+m}} \frac{\ds d\;}{\ds d\alpha_j}\right) \;{\cal G}_o((\alpha,p)_{n+m}){\cal G}_{n,m}((\alpha,p)_{n+m})\end{array} \end{equation}when $(\alpha)_N=0$. The generators (\ref{free-gen}) and (\ref{genr2}) exhibit the appropriate factorizations.\begin{equation}\label{genfactr} \renewcommand{\arraystretch}{1.25} \begin{array}{l} {\cal K}_{n,m}((\alpha,-p)_{n,1},(\beta,q)_m)=\ln{\cal G}_o((\alpha,-p)_{n,1},(\beta,q)_m)+\ln{\cal G}_{n,m}((\alpha,-p)_{n,1},(\beta,q)_m)\\
 \qquad\qquad = {\ds \int} {\ds \frac{d{\bf u}\quad}{(2\pi)^3}}\; \left( {\ds \sum_{k=1}^n} e^{\overline{i{\bf p}_k\cdot{\bf u}}}\sqrt{2\omega_k}\,\delta^+(p_k)\alpha_k \right) \left( {\ds \sum_{j=1}^m} e^{i{\bf q}_j\cdot{\bf u}}\sqrt{2\omega_{\hat{j}}}\,\delta^+(q_j)\beta_j \right)\\
 \qquad\qquad\qquad + {\ds \int d\sigma(\lambda)\int}{\ds \frac{du}{(2\pi)^4}}\; {\ds \prod_{k=1}^n} \left( a_k\! + \!\lambda e^{\overline{ip_j u}} \hat{\delta}(p_k) \alpha_k\right) {\ds \prod_{j=1}^m} \left( a_j\! + \!\lambda e^{iq_j u} \hat{\delta}(q_j)\beta_j \right).\end{array}\end{equation}Derived from the $a_{kj}$ in (\ref{genr2}), $a_k=0$ for $k=1,2$ and $a_k=1$ otherwise. After the argument relabeling, $a_{kj}=0$ corresponds with $n,m=1,2$ in $\exp({\cal K}_{n,m})$. The ${\cal K}_{n,m}((\alpha,-p)_{n,1},(\beta,q)_m)$ defined by the construction (\ref{w-defn}) is in the form of (\ref{t-factor}). $\zeta={\bf u}$ for the first contribution to the summation with\[A_n({\bf u},(\alpha,p)_n)= {\ds \sum_{j=1}^n} e^{i{\bf p}_j\cdot{\bf u}}\sqrt{2\omega_{\hat{j}}}\,\delta^+(p_j)\alpha_j\]and $\zeta = u,\lambda$ for a second contribution with\[A_n(\lambda,u,(\alpha,p)_n)={\ds \prod_{j=1}^n} \left( a_j\! + \!\lambda e^{ip_j u} \hat{\delta}(p_j)\alpha_j \right).\]Both measures in (\ref{genfactr}) are nonnegative. Using T.5, this factorization (\ref{genfactr}) with the nonnegative weight for ${\cal K}_{n,m}((\alpha,p)_{n+m})$ demonstrates that the semi-norm (\ref{norm}) applies in ${\cal B}$ and A5 is satisfied.

\subsection{Continuous linear functional}\label{sec-linf}
Axiom A1 asserts that $\underline{W}$ is a continuous linear functional dual to the algebra ${\cal A}$, that ${\cal A}$ includes functions of bounded spacetime support and that the subalgebra ${\cal B}$ includes functions with Fourier transforms of bounded energy-momentum support. From sections \ref{sec-A} and \ref{sec-B}, the support conditions are satisfied. The demonstration now verifies that the construction (\ref{w-defn}) defines continuous linear functionals.
\newline

\addtocounter{theorems}{1}
{\em Theorem} \thetheorems: The $W_n((x)_n)$ defined in (\ref{w-defn}) are continuous linear functionals dual to ${\cal A}$ when the number of spacetime dimensions is three or more and $m>0$. 

From (\ref{w-defn}) using (\ref{free-gen}), (\ref{genr2}) and (\ref{faadibruno}), each term in a $W_n((x)_n)$ are products of factors of conjoined functions with no arguments in common. As a consequence, if the conjoined functions defined by (\ref{twopoint}) and (\ref{c-defn}) are continuous linear functionals, then the $W_n((x)_n)$ are continuous linear functionals. The two-point function of a free field (\ref{twopoint}) is well-defined. The issue is whether the higher order conjoined functions (\ref{c-defn}) are generalized functions. From the definitions in section \ref{sec-A}, the ${^nC}_{k,\eta}(x_{i_1},\ldots x_{i_\eta})$ from (\ref{c-defn}) are continuous linear functionals if the implied $\tilde{T}_{(s)_n}(({\bf p})_n)$ in (\ref{emsupp}) are elements of ${\cal S}'({\bf R}^{3n})$. In this section, the number of spacetime dimensions is considered and is designated as $d$.\begin{equation} \tilde{T}_{(s)_n}(({\bf p})_n) =\prod_{j=1}^n \frac{1}{2\omega_j}\; \delta(\omega_1 \ldots +\omega_k -\omega_{k+1} \ldots -\omega_n)\;\delta({\bf p}_1\!+\!{\bf p}_2\ldots \!+\!{\bf p}_n).\label{delta2}\end{equation}Here, for notational convenience, the negative energies are designated $1$ through $k$ and the positive energies by $k+1$ through $n$. That is, $s_j=1$ for $j\leq k$ and $s_j=-1$ otherwise. $\tilde{W}_n((p)_n)$ includes permutations of this assignment. Factors of $1/(2\omega_j)$ are multipliers of ${\cal S}({\bf R}^3)$ and are not considered further.

$\delta({\bf p}_1\!+\!{\bf p}_2\ldots \!+\!{\bf p}_n)$ constrains the evaluation of the generalized function (\ref{delta2}) to summation on the subsurface of $({\bf p})_n \in {\bf R}^{3n}$ with momentum conserved,\[{\bf p}_n=-{\bf p}_1\ldots -{\bf p}_{n-1}.\]Within this subsurface, summation over the subsurface with the infinitely differentiable\begin{equation}\label{subsurface}E_k(({\bf p})_n):=\sum_{j=1}^n s_j \omega_j=0\end{equation}defines a generalized function except possibly for points on the subsurface with a vanishing gradient [\ref{gel1}]. The result demonstrated here is that the divergence due to vanishing of the gradient on the subsurface with $E_k(({\bf p})_n)=0$ is summable when $d\geq 3$. [\ref{gel1}] includes the demonstration that an infinitely differentiable function $E_k(({\bf p})_n)$ defines a generalized function $\delta(E_k(({\bf p})_n))$ except where the gradient vanishes on the subsurface $E_k(({\bf p})_n)=0$. The gradient of the function $E_k(({\bf p})_n)$ from (\ref{subsurface}) does vanish when $E_k(({\bf p})_n)=0$ in limited cases and an exploration of the generalized function for those cases is developed below. An analogous case is the divergence of $\delta(r^k)$ except when summed in $k$ or more dimensions when $r$ is the Euclidean distance from the origin.

The components of the gradient of $E_k(({\bf p})_n)$ on the subsurface with momentum conserved are\begin{equation}\label{grad} \renewcommand{\arraystretch}{2.25} \begin{array}{rl} {\ds \frac{dE_k(({\bf p})_n)}{dp_{j(\ell)}}} &= s_j {\ds \frac{d\omega_j\;}{dp_{j(\ell)}}}+s_n {\ds \frac{d\omega_n\;}{dp_{n(\ell)}}}\, {\ds \frac{dp_{n(\ell)}}{dp_{j(\ell)}}}\\
 &= s_j \frac{\ds p_{j(\ell)}}{\ds \omega_j}-s_n\frac{\ds p_{n(\ell)}}{\ds \omega_n}\end{array}\end{equation}from (\ref{subsurface}) with momentum coordinates ${\bf p}_j$ labeled $p_{j(1)},p_{j(2)},\ldots p_{j(d-1)}$ and $j=1$ through $n-1$. Summing squares provides that when the gradient vanishes, $\omega_j=\omega_n$ for any $j$. Then, the gradient vanishes if and only if $s_j{\bf p}_j=s_n{\bf p}_n$ for each $j$. Only the cases with $s_1=1$ and $s_n=-1$ need be considered since $E_0\neq 0$ and $E_n\neq 0$.
 
A neighborhood $V$ of those points with a vanishing gradient is given by\[{\bf p}_j =s_j {\bf p}_1 +{\bf e}_j\]for $j \in \{2,n-1\}$ and $\| {\bf e}_j \| < \epsilon$. In $V$,\begin{equation}\label{pn-defn}{\bf p}_n=-\sum_{j=1}^{n-1}{\bf p}_j=(n-1-2k){\bf p}_1-\sum_{j=2}^{n-1} {\bf e}_j\end{equation}and for $j \in \{2,n-1\}$,\begin{equation}\label{expnd-w}\omega_j \approx \omega_1 + s_j \frac{{\bf p}_1\cdot {\bf e}_j}{\omega_1} +\frac{{\bf e}_j \cdot {\bf e}_j}{2\omega_1}-\frac{({\bf p}_1 \cdot {\bf e}_j)^2}{2\omega_1^3}\end{equation}to second order in small quantities. From $\omega_j=\omega_n$ for each $j\in \{1,\ldots n-1\}$, when the gradient vanishes, it follows from (\ref{subsurface}) that $E_k(({\bf p})_n)=(n-2k)\omega_n$ when the gradient vanishes. As a consequence, the gradient vanishes when $E_k(({\bf p})_n)=0$ if and only if $2k=n$. When $2k=n$ and within $V$, ${\bf p}_n =-{\bf p}_1 +{\bf e}_n$ with\[{\bf e}_n := -\sum_{\ell=2}^{n-1} {\bf e}_{\ell}\]from (\ref{pn-defn}).

The only simultaneous solutions to $E_k(({\bf p})_n)=0$ and grad$\,E_k(({\bf p})_n)=0$ on the subsurface with momentum conserved have $({\bf e})_{2,n\!-\!1}=0$ and $2k=n$. Within $V$ for this singular case,\[\renewcommand{\arraystretch}{1.25} \begin{array}{rl} E_k(({\bf p})_n) ={\ds \sum_{j=1}^n} s_j \omega_j &\approx \frac{\ds 1}{\ds 2\omega_1^3}\; {\ds \sum_{j=2}^n} s_j \left( \omega_1^2\; {\bf e}_j\cdot{\bf e}_j-({\bf p}_1 \cdot {\bf e}_j)^2\right)\\
 &:= \frac{\ds R^2}{\ds 2\omega_1^3}\;(a{\bf p}_1^2 +bm^2) \end{array}\]with\[a R^2:= {\ds \sum_{j=2}^n} s_j \left( {\bf e}_j\cdot{\bf e}_j-({\bf u}_1 \cdot {\bf e}_j)^2\right),\qquad\qquad
b R^2:= {\ds \sum_{j=2}^n} s_j \, {\bf e}_j\cdot{\bf e}_j,\]${\bf u}_1 :={\bf p}_1/ \|{\bf p}_1\|$, the unit vector in the direction of ${\bf p}_1$, and polar coordinates for $({\bf e})_{2,n-1}$.\[R^2=\sum_{j=2}^{n-1} {\bf e}_j\cdot{\bf e}_j.\]Variations of $a,b$ with $R$ are negligible in the small neighborhood $V$, $\epsilon \ll 1$.
 
The form (\ref{delta2}) has $1/R^2$ and $\delta(R^2)$ divergences within $V$. When $2k=n$ and within $V$,\[\renewcommand{\arraystretch}{2.25} \begin{array}{rl} \delta(E_k(({\bf p})_n)) &= \delta(\frac{\ds R^2}{\ds 2\omega_1^3}\;(a{\bf p}_1^2 +bm^2) )\\
 &= \frac{\ds 2\omega_1^3}{\ds R^2}\;\delta(a{\bf p}_1^2 +bm^2) + \frac{\ds 2\omega_1^3}{\ds a{\bf p}_1^2 +bm^2}\; \delta(R^2 ).\end{array}\]$\delta(a{\bf p}_1^2 +bm^2)$ is defined since $a,b$ are independent of $R$ and the generalized function is defined when the gradient is finite, $R>0$. Then, (\ref{delta2}) defines a generalized function since the singularities are locally summable. $d\geq 3$ suffices since the Jacobian for the polar coordinates for $({\bf e})_{2,n-1}$ contributes $R^{(d-1)(n-2)-1}$ and $n\geq 4$ for the conjoined functions (\ref{c-defn}). The $R^{(d-1)(n-2)-1} \delta(R^2)$ term vanishes for $d\geq 3$ from $x\delta(x)=0$.

Then (\ref{w-defn}) describes generalized functions dual to ${\cal A}$. This completes the demonstration that the construction (\ref{w-defn}) satisfies the axioms A1-A5 for the function sequences ${\cal A}$ from section \ref{sec-A} and ${\cal B}$ from section \ref{sec-B}. This is the primary result of the study and the results are summarized by:
\newline

{\em Main Theorem}: The sequence $\underline{W}$ of $n$-point generalized functions $W_n((x)_n)$ defined in (\ref{w-defn}) satisfy axioms A1-A5 for the function sequences ${\cal A}$ and ${\cal B}$.

The representation of the Poincar\'{e} group supports the conclusion that the $W_n((x)_n)$ realize a QFT for a single, neutral, Lorentz scalar field. Interaction is demonstrated in section \ref{sec-satisfy}. 

\section{Properties of the constructions}\label{sec-satisfy}
In this section, additional properties of the constructions are developed: the constructions include interaction as exhibited in plane wave limit scattering amplitudes when $c_n>0$; the vacuum is in a one-dimensional subspace of translational-invariant states; there are no elements of ${\cal B}$ with spatially bounded support; vacuum polarization can be implemented; and the generator of time translation coincides with the Hamiltonian described in canonical quantizations as a free field Hamiltonian although the constructions exhibit interaction.

\subsection{Interaction}\label{sec-scat}
\addtocounter{theorems}{1}
{\em Theorem} \thetheorems: The plane wave limit scattering amplitudes exhibit interaction.

Scattering amplitudes are large time difference limits of state transition amplitudes. The LSZ (Lehmann-Symanzik-Zimmermann) expressions [\ref{bogo}] for scattering amplitudes are\begin{equation}\label{S-defn}S:={\ds \lim_{t \rightarrow \infty}} \langle U(t) \underline{\ell}(t) | U(-t) \underline{\ell}'(-t) \rangle\end{equation}for sequences of LSZ functions $\underline{\ell}(t)$ and $\underline{\ell}'(t)$, and time translation $U(t)$. $U(t)$ is the unitary realization of time translation that results from translation invariance of the Wightman-functional [\ref{pct}].
\newline

\addtocounter{defintn}{+1}
{\em Definition} \thedefintn: The Fourier transforms of the LSZ functions are\begin{equation}\label{LSZ}\tilde{\ell}_n(\tau):=\tilde{\ell_n}(\tau;(p)_n)=\prod_{j=1}^n \;e^{i\omega_j \tau}(\omega_j + E_j) \tilde{f}(({\bf p})_n)\end{equation}with $\tilde{f}(({\bf p})_n)$ a Schwartz tempered test function and $\tau$ a real parameter. 

A result of (\ref{a-struct}) and (\ref{B-defn}), these LSZ functions are elements of ${\cal B}$. A number of familiar forms derive from the definition. For any $n$-point generalized function argument,\begin{equation}\label{LSZ-spt}\renewcommand{\arraystretch}{1.75} \begin{array}{rl} U(t)\Phi(\ell_1(t))U(t)^{-1} &= i {\ds \int} d{\bf x}\; \hat{u}(t,{\bf x}) \stackrel{\leftrightarrow}{\partial}_o \Phi(t,{\bf x})\\
 &= {\ds \int} dp\; (\omega+E) e^{i(\omega-E)t}\tilde{f}({\bf p})\;\tilde{\Phi}(p)\\
 &=\Phi(\ell_1(0))\end{array}\end{equation}with $f(x) \stackrel{\leftrightarrow}{\partial}_o g(x):= f(x)\dot{g}(x)-\dot{f}(x)g(x)$, $\dot{f}(x)$ the first time derivative of $f(x)$ and\[\hat{u}(x) :=\frac{1}{(2\pi)^2}\int d{\bf p}\; e^{i\omega t}e^{-i{\bf p}\cdot {\bf x}}\tilde{f}({\bf p})\]is a smooth solution of the Klein-Gordon equation. $\Phi(x_j)$ indicates the $j$th argument of the functional $W_n((x)_n)$ using (\ref{sesquis2}) and $U(\tau)\Phi(t,{\bf x})U(\tau)^{-1}=\Phi(t+\tau,{\bf x})$ indicates that the corresponding argument of the functional $W_n((x)_n)$ is translated in time by $\tau$. $U(t)\Phi(\ell_1(t))U(t)^{-1}$ is independent of $t$ due to the concentration of the support of the $\tilde{W}_n((p)_n)$ within mass shells and the time-dependent form of the LSZ functions (\ref{LSZ}).

A convenient selection for evaluation of plane wave limits of the scattering amplitudes is\begin{equation}\label{LSZdelta}\tilde{f}({\bf p}) =\left(\frac{L}{\sqrt{\pi}} \right)^3 e^{-L^2 ({\bf p} -{\bf q})^2}>0,\end{equation}a point-wise nonnegative delta sequence of Schwartz tempered test functions. Contributions are heavily weighted near the momentum ${\bf q}$ in the plane wave limit as $L$ grows without bound. The scattering amplitudes are evaluated for plane wave ``in'' states\begin{equation}\label{instates} \lim_{\stackrel{L \rightarrow \infty}{t\rightarrow -\infty}} |U(t) \tilde{\ell}_n(t) \rangle \rightarrow |(q)_n^{\mathit{in}} \rangle \end{equation}with ``out'' states the $t\rightarrow \infty$ limits. The relation between this normalization (\ref{instates}) for $|(q)_n^{\mathit{in}} \rangle$ and the box normalization common in Feynman series is developed in [\ref{feymns}] and uses theorem T.7, that the $W_n((x)_n)$ coincide with the free field functions when $c_n=0$.

Using (\ref{S-defn}), (\ref{LSZ}) and (\ref{LSZdelta}), the scattering amplitudes are limits of quadratures. In the notation (\ref{instates}) and from (\ref{b-ident1}) and (\ref{LSZ-spt}), the non-forward scattering amplitudes are\[\renewcommand{\arraystretch}{1.25} \begin{array}{rl} \langle (q)_n^{\mathit{in}}|(q)_{n+1,n+m}^{\mathit{out}} \rangle &= {\ds \lim_{L\rightarrow \infty}} {^CW}_{n+m}(\ell_n(0)^* \ell_m(0))\\
 &= {\ds \lim_{L\rightarrow \infty}} c_{n+m}\;\left( {\ds \frac{L}{\sqrt{\pi}}} 
\right)^{3(n+m)}{\ds \int} d({\bf p})_{n+m} \; {\ds \prod_{j=1}^{n+m}}\; e^{-L^2 ({\bf p}_j-{\bf q}_j)^2} \times \\
 &\qquad \delta(\omega_1 \ldots \!+\!\omega_n\!-\!\omega_{n+1} \ldots \!-\!\omega_{n+m})\;\delta({\bf p}_1 \ldots \!+\!{\bf p}_n\!-\!{\bf p}_{n+1} \ldots \!-\!{\bf p}_{n+m}).\end{array}\]The definition of the $*$-mapped function (\ref{mapf}), relabeling of the first $n$ momentum summation variables ${\bf p}_j\mapsto -{\bf p}_j$, the symmetry of (\ref{c-defn}) and evaluation of the mass shell deltas simplify the expression.

This quadrature is readily evaluated in the plane wave limit. With the contribution to the summation heavily weighted near $({\bf p})_{n+m}=({\bf q})_{n+m}$ and to leading order in small differences, Taylor expansion results in\[\omega_1 \ldots \!+\!\omega_n\!-\!\omega_{n+1} \ldots \!-\!\omega_{n+m} \approx \sum_{k=1}^{n+m} (s_k\omega({\bf q}_k)+s_k{\bf b}_k\cdot({\bf p}_k-{\bf q}_k))\]with\[{\bf b}_k:=\frac{{\bf q}_k}{\omega({\bf q}_k)},\]$s_j:=1$ for $1\leq j \leq n$ and $s_j:=-1$ for $n<j\leq n+m$. With $u:=\upsilon,{\bf u}$ and using the Fourier transform of $\delta(x)$,\[\renewcommand{\arraystretch}{1.25} \begin{array}{l} {^CW}_{n+m}(\ell_n(0)^* \ell_m(0))\approx c_{n+m}\; \left( {\ds \frac{L}{\sqrt{\pi}}} \right)^{3(n+m)} {\ds \int} {\ds \frac{du}{(2\pi)^4}} \; {\ds \int} d({\bf p})_{n+m} \; \times \\
 \qquad \qquad {\ds \prod_{j=1}^{n+m}}\; e^{is_j(\omega({\bf q}_j)+{\bf b}_j\cdot({\bf p}_j-{\bf q}_j))\upsilon} e^{is_j{\bf p}_j \cdot {\bf u}}e^{-L^2 ({\bf p}_j-{\bf q}_j)^2}\\
 \qquad = c_{n+m}\; \left( {\ds \frac{L}{\sqrt{\pi}}} \right)^{3(n+m)} {\ds \int} {\ds \frac{du}{(2\pi)^4}} \; {\ds \int} d({\bf p})_{n+m} \; {\ds \prod_{j=1}^{n+m}}\; e^{is_j(\omega({\bf q}_j)+{\bf b}_j\cdot{\bf p}_j)\upsilon} e^{is_j({\bf p}_j+{\bf q}_j)\cdot {\bf u}}e^{-L^2 {\bf p}_j^2}\end{array}\]after relabeling each summation for a translation of ${\bf p}_k$ by ${\bf q}_k$. From the development in section \ref{sec-linf}, the leading order from the Taylor expansion contributes for a non-forward selection of $({\bf q})_{n+m}$. The remaining summations are elementary using\[\sqrt{\alpha} \int_{-\infty}^{\infty} ds\; e^{-\alpha s^2 +\beta s} = \sqrt{\pi} \; e^{\beta^2/(4 \alpha)}.\]With\[{\bf q}:=\sum_{j=1}^{n+m} s_j {\bf q}_j\quad \mbox{and}\qquad q_{0}:=\sum_{j=1}^{n+m} s_j \omega({\bf q}_j)\]and from $s_j^2=1$,\[\renewcommand{\arraystretch}{1.25} \begin{array}{l} {^CW}_{n+m}(\ell_n(0)^* \ell_m(0))= c_{n+m}\; {\ds \int} {\ds \frac{du}{(2\pi)^4}}\; e^{iq_0 \upsilon+i{\bf q}\cdot {\bf u}} {\ds \prod_{j=1}^{n+m}}\;e^{-({\bf b}_j\upsilon +{\bf u})^2/(4L^2)}\\
 \qquad = c_{n+m}\; {\ds \int} {\ds \frac{du}{(2\pi)^4}}\; e^{iq_0 \upsilon+i{\bf q}\cdot {\bf u}} \;e^{-(n+m)(({\bf b}_s^2-{\bf b}^2)\upsilon^2+({\bf u}+{\bf b}\upsilon)^2)/(4L^2)}\\
 \qquad = c_{n+m} \left({\ds \frac{L}{\sqrt{\pi (n+m)}}}\right)^4 \; {\ds e^{-L^2 {\bf q}^2/(n+m)}
}\; {\ds \frac{ {\ds e^{-L^2 (q_0-{\bf q}\cdot{\bf b})^2/((n+m)\sigma_b^2)}}}{\sigma_b}}\end{array}\]with\[{\bf b}:=\frac{1}{n+m}\sum_{j=1}^{n+m} {\bf b}_j,\qquad\qquad {\bf b}_s^2:=\frac{1}{n+m}\sum_{j=1}^{n+m} {\bf b}_j^2\]and\[\sigma_b^2:={\bf b}_s^2-{\bf b}^2=\frac{1}{(n+m)^2}\sum_{k=1}^{n+m}\sum_{j=k+1}^{n+m}({\bf b}_k-{\bf b}_j)^2>0.\]Finally, the large $L$ limit for the non-forward contribution to the scattering amplitude is\begin{equation}\label{scaty}\langle (q)_n^{\mathit{in}}|(q)_{n+1,n+m}^{\mathit{out}} \rangle = c_{n+m}\;\delta(q_1 \ldots\!+\!q_n \!-\!q_{n+1}\ldots \!-\!q_{n+m})\end{equation}and the energies of the $q_j$ are all on positive mass shells.

Demonstration that the limit (\ref{S-defn}) is nontrivial for the choice of LSZ functions (\ref{LSZdelta}) from ${\cal B}$ suffices to demonstrate interaction.

\subsection{Uniqueness of the vacuum}\label{sec-vac}
The first result is that the $n$-point connected functions are connected. When $f_k((x)_k)$ and $f_{n-k}((x)_{n-k}) \in {\cal B}$, the development of (\ref{b-ident1}) applied to (\ref{connctd}) results in\[{^CW}_n(f_k^*f_{n-k})={^nC}_{k,n}\left(\hat{\bf S}[f_k^*]\,\hat{\bf S}[f_{n-k}]\right).\]Symmetry with transpositions of arguments in (\ref{c-defn}) results in the useful identity\begin{equation}\label{wc-defn}\begin{array}{l} {^CW}_n(f_k^* f_{n-k})={^nC}_{k,n}\left(f_k^*f_{n-k}\right)\\
 \qquad \qquad =c_n {\ds \int} d(p)_n\;\delta(p_1+\ldots p_n)\,{\ds \prod_{j=1}^k}\delta^-(p_j)\,\widetilde{f^*}_k((p)_k){\ds \prod_{\ell=k+1}^n}\delta^+(p_\ell) \tilde{f}_{n-k}((p)_{k+1,n})\end{array}\end{equation}when $k\neq 0,1,n-1,n$, and\begin{equation}\label{wn-zero}{^CW}_n(f_n)={^CW}_n(f_1^* f_{n-1})=0.\end{equation}

\addtocounter{theorems}{1}
{\em Lemma} \thetheorems: The conjoined functions ${^nC}_{k,n}((x)_n)$ are connected.

From (\ref{poin-trans}), (\ref{wc-defn}) and the Riemann-Lebesgue lemma [\ref{titchmarsh}],\begin{equation}\label{decline}{^nC}_{k,n}(f_k^*\,(\rho a,1) f_{n-k})\rightarrow 0\end{equation}for $1\leq k <n$ and $a^2<0$ as the real number $\rho$ grows without bound when $f_k((x)_k)$ and $f_{n-k}((x)_{n-k}) \in {\cal B}$ [\ref{gej05}]. From section \ref{sec-linf}, the generalized functions of momenta $\tilde{T}_n({\bf p})_n)$ defined using (\ref{emsupp}) for (\ref{wc-defn}) are locally absolutely summable functions of $({\bf p})_n$. As a consequence, evaluation of (\ref{wc-defn}) results in a summable function of $({\bf p})_n$ that includes a factor $\exp(i\rho\, {\bf a}\cdot ({\bf p}_1+\ldots {\bf p}_k))$ with $1<k<n-1$. The Riemann-Lebesgue lemma provides the result (\ref{decline}) and the symmetry of the conjoined functions (\ref{c-defn}) with transpositions of arguments results in the lemma.
\newline

\addtocounter{theorems}{1}
{\em Theorem} \thetheorems: For Wightman-functionals defined by (\ref{w-defn}), the subspace of translational-invariant states is one-dimensional.

From (\ref{wc-defn}),\begin{equation}\label{decline2}{^CW}_n(f_k^*\,(\rho a,1) f_{n-k})\rightarrow 0\end{equation}for the same conditions as (\ref{decline}). $W_0(f_0)=f_0$ and $W_1=0$ from section \ref{sec-constr}. $W_n(f_n)=0$ for $f_n((x)_n)\in {\cal B}$ and $n> 1$ from (\ref{twopoint}) and (\ref{wn-zero}). As a consequence of (\ref{decline2}) and the evaluations of $W_n(f_n)$,\begin{equation}\label{cl-decomp}\underline{W}(\underline{f}^*\,{\bf x}\, (\rho a,1) \underline{g})=f_0g_0 = \underline{W}(\underline{f}^*)\underline{W}(\underline{g})\end{equation}for $\underline{f},\underline{g} \in {\cal B}$ as $\rho$ grows without bound for $a^2<0$.

This cluster decomposition (\ref{cl-decomp}) implies that the subspace of translational-invariant states is one-dimensional [\ref{borchers}]. The equivalence class of $\underline{1}$ is one translational-invariant state and is designated the vacuum $\Omega$. If $\underline{f}\in {\cal B}$ labels a second, linearly independent translational-invariant state in the Hilbert space, then $\underline{f}$ can be selected with $\langle \Omega |\underline{f}\rangle=0$ by Gram-Schmidt construction. From the translational invariance,\[\renewcommand{\arraystretch}{1.25} \begin{array}{rl} \underline{W}(\underline{f}^*\,{\bf x}\,\underline{f})&=\underline{W}(\underline{f}^*\,{\bf x}\, (\rho a,1) \underline{f})\\
 &= \underline{W}(\underline{f}^*)\underline{W}(\underline{f})\\
 &=|\underline{W}(\underline{1}^*\,{\bf x}\, \underline{f})|^2\\
 &=|\langle \Omega |\underline{f}\rangle|^2\\
 &=0\end{array}\]from the isometry (\ref{isometry}) and then $\|\underline{f}\|_{\cal B} =0$ in contradiction to the assertion that $\underline{f}$ labels a second, linearly independent translational-invariant state.

\subsection{${\cal B}$ lacks functions of bounded spatial support}\label{sec-b-supprt}
\addtocounter{theorems}{1}
{\em Theorem} \thetheorems: There are no elements within ${\cal B}$ of bounded spatial support. For an $f(x)\in {\cal B}$, there is no ${\bf x}_o\in {\bf R}^3$ and finite $R$ such that $f(x)=0$ for $\|{\bf x}-{\bf x}_o\|>R$.

If there is an $f(x)\in {\cal B}$ such that $f(x)=0$ for all ${\bf x}\in U \subset {\bf R}^3$, then the Fourier transform\[u(E,{\bf x}):=2\pi\,\int dt\; e^{-itE} f(x) =0\]for ${\bf x}\in U$ and any $E$. $U:=\{{\bf x} ; \; \|{\bf x}-{\bf y}\|<\rho\,\}$ for a finite $\rho$, a selected ${\bf y}\in {\bf R}^3$ and $u(E,{\bf x})$ is infinitely continuously differentiable from (\ref{a-struct}) and (\ref{B-defn}). The Fourier transform of a Schwartz function is a Schwartz function [\ref{gel1}]. From (\ref{B-defn}),\[\renewcommand{\arraystretch}{1.25} \begin{array}{rl} u(E,{\bf x})&:= {\ds \int} d{\bf p}\; e^{-i{\bf px}}(E+\omega)\tilde{\varphi}(E,{\bf p})\\
 &=(E +(m^2-\Delta)^{\frac{1}{2}})\, g(E,{\bf x})\end{array}\]with $\varphi(x)\in {\cal A}$, $\Delta$ the Laplacian and\[ g(E,{\bf x}):= \int d{\bf p}\; e^{-i{\bf px}}\tilde{\varphi}(p).\]Linear independence provides that when $u(E,{\bf x})=0$ for any $E$, that\[g(E,{\bf x})=(m^2-\Delta)^{\frac{1}{2}} \, g(E,{\bf x})=0.\]As a consequence, when $f(x)=0$ for all ${\bf x}\in U$, then $g=(m^2-\nabla)^{\frac{1}{2}}\, g=0$ within the finite region ${\bf U}$ for any $E$ and [\ref{reeh},\ref{segal},\ref{masuda}] provides that $g=0$. $g=0$ implies that $f(x)=0$. Then, the only $f(x)\in {\cal B}$ that vanishes for ${\bf x}\in U$ is $f(x)=0$ and $f(x)=0$ is the only element of ${\cal B}$ that vanishes everywhere outside a bounded spatial region. This development applies to each argument of an $f_n((x)_n)\in {\cal B}$.

\subsection{Vacuum polarization}
$W_1=0$ for the constructions defined by (\ref{free-gen}) and (\ref{genr2}). A real constant vacuum polarization $W_1$ can be included as a character [\ref{heger}] and implemented by an additional generator factor $\exp(\sum_{k=1}^n \alpha_k W_1)$. This generator factors appropriately to preserve the semi-norm A5. $W_1\neq 0$ adds $p_j=0$ to the supports of Fourier transforms of the $W_n((x)_n)$ but validity of A1-A4 remains without additional modifications.

\subsection{The Hamiltonian}\label{sec-hamil}
The concentration of the energy-momentum support of the $\tilde{W}_n((p)_n)$ within mass shells and the zeros on negative energies of functions in ${\cal B}$ provide that evolution with time is analogous to the time evolution of a free field.\begin{equation}\label{hamil} \langle \underline{g}| U(t) \underline{h}\rangle = {\ds \sum_{n,m} \int} d(p)_{n+m}\; \overline{\tilde{g}_n((-p)_{n,1})}\;\tilde{W}_n((p)_{n+m})\; {\ds \prod_{k=n+1}^{n+m}} e^{-i\omega_k t}\;\tilde{h}_m((p)_{n+1,n+m})\end{equation}for $\underline{g}, \underline{h} \in {\cal B}$ and the unitary time translation $U(t)$.

\section{Conclusions}\label{conclusion}
\subsection{Decisive revision}\label{conclu}
UQFT differ substantially from prior developments and conjecture for QFT. In addition to differences developed in the discussion above, differences include:\begin{enumerate}\item In UQFT, nontrivial interaction is consistent with a two-point function in the form of a free field two-point function, the positive frequency Pauli-Jordan function. From section \ref{sec-constr}, $W_2(x_1,x_2)=\Delta(x_1,x_2)$ in ${\cal B}$. The Jost-Schroer theorem (lemma 21.1 [\ref{bogo}]) and similar results [\ref{feder},\ref{greenberg}], that $n$-point generalized functions are determined as free field VEV (\ref{wodefn-free}) when the two-point function is the Pauli-Jordan function, does not result from the revised spectral support condition A3. Assuming for argument that Hilbert space field operators are defined, application of A3 does not imply that $\Phi^-(f)\Omega=0$ nor $\Phi^-(f)\Phi^+(g)\Omega =c\; \Omega$ unless $f\in {\cal B}$ and then $c=0$. $\Omega=|\underline{1}\rangle$ is the vacuum state and the field is decomposed into positive and negative energy components, $\Phi(f)=\Phi^+(f)+\Phi^-(f)$. $\Phi^-(g)=0$ when $g\in {\cal B}$ and $\Phi^-(g^*)$ maps elements out of the Hilbert space based upon ${\cal B}$. A conclusion that the commutator $[\Phi^-(x),\Phi^+(y)]$ is the product of the positive frequency Pauli-Jordan function and the identity operator does not result from A1-A5. Indeed,\[W_n(g_1^*\ldots g_k^* g_{k+1}\ldots g_n)=\langle \Omega| \prod_{\ell=1}^k \Phi^-(g_\ell^*)\prod_{j=k+1}^n \Phi^+(g_j)\Omega\rangle\]for the constructions and symmetry of the generalized functions (\ref{w-defn}) provides that the operators $\Phi^\pm (f)$ commute. The number and type of factors $\Phi^\pm(f)$ with $f=g$ or $f=g^*$, $g\in{\cal B}$ determines the scalar product (\ref{sesquis}). And finally, the $n$-point generalized functions are not determined by the two-point function when Hilbert space field operators are lacking.

In the case of the free field, the components ${\underline{W}\!\downharpoonright}_{\,\cal B}$ of the Wightman-functional that contribute in ${\cal B}$ are augmented with $n$-point generalized function terms to define the free field Wightman-functional for ${\cal A}$, ${\underline{W}\!\downharpoonright}_{\,\cal B}\;\mapsto \underline{W_{\cal A}} \neq \underline{W}$. The field operators resulting from $\underline{W_{\cal A}}$ have the conventional commutation relations and $\Phi^-(f)\Omega=0$ for $f\in {\cal A}$. 
 
\item A consequence of the zeros of the Fourier transforms of functions in ${\cal B}$ is that the functions labeling the elements of the Hilbert space do not vanish in any spatial neighborhood when interaction is present. Then, narrowing consideration of observables to orthogonal projection operators onto subspaces of states, an association of observables with bounded subsets of spacetime can only be approximate. From theorem T.13, the elements in ${\cal B}$ with bounded temporal support can not have bounded spatial support. In the case of the free field, an extension of the semi-norm from ${\cal B}$ to the set ${\cal A}$ of functions that include functions of bounded spacetime support preserves satisfaction of the Wightman axioms, and the isotony condition of the Haag-Kastler algebraic QFT description applies to the projection operators as observables. This extension does not apply when interaction is exhibited. In this sense, free fields are singularly removed from fields exhibiting interaction.\end{enumerate}

\subsection{Summary}
This study relies only on general principles of relativistic quantum physics to construct realizations of Wightman-functionals that exhibit interaction in physical spacetime. An appropriate Hilbert space of positive energy states, Poincar\'{e} covariance and local commutativity are considered more strongly motivated conditions than an involutive algebra of sequences of Schwartz functions and the consequent Hermitian Hilbert space field operators. Interest in these constructions includes that they are explicit and peculiarly quantum mechanical alternatives to canonical quantization of classical field equations.
 
Fields are classical dynamic quantities and association of quantum fields with Hermitian Hilbert space field operators has been an assertion used to derive axioms for relativistic quantum physics as well as the Feynman rules. However, self-adjointness of the field is a conjecture that substantially limits a functional analytic development of relativistic quantum physics. In the constructions, quantum fields are defined conventionally as a multiplication in the algebra of function sequences but these fields necessarily are not Hermitian Hilbert space field operators when interaction is present. Free fields and related physically trivial cases are exceptional. Lacking Hermitian Hilbert space field operators, the demand that the description of the quantum field is a canonical quantization of a classical description is relaxed. Indeed, for the constructions, the interaction Hamiltonians vanish although interaction is manifest. Peculiarly quantum mechanical descriptions with alternative classical limits result. The constructed, explicit Wightman-functionals evaluated for selected states describe classical limits although a general correspondence of quantum with classical fields is yet to be determined. It is anticipated that descriptions of UQFT are determined in part by the classical limits, but in associations other than canonical quantizations [\ref{pct}]. Classical limits of quantum mechanics are richer than the conflation of observable with self-adjoint operator [\ref{johnson}].

Descriptions of measurement processes and interpretations of quantum physics, controversial subjects since the inception of quantum mechanics, have a substantial impact on whether these constructions that lack self-adjoint field operators are considered. Indeed, while there is a negligible physical difference between:\begin{enumerate}\item measurement results from the collapse of a state to an eigenstate of the observable that is necessarily realized as a self-adjoint operator in the Hilbert space, and \item measurement results from entanglement of states of the observer with states of the observed system that have approximately the same physical descriptions as eigenstates of the observable,\end{enumerate}the mathematical distinction is decisive. The ${\bf x^3p}$ and ${\bf x}$ examples illustrate that only the second of these two descriptions generally applies in the rigged Hilbert spaces of quantum physics. The first description, while applicable in finite dimensional Hilbert spaces, is not generally realizable. Indeed, not every Hilbert space supports implementation of particular self-adjoint operators nor their eigenstates but projections onto subspaces of states are inherent to every Hilbert space. And, a state collapse incurs the Einstein-Podolsky-Rosen and Schr\"{o}dinger's cat measurement paradoxes.


\section*{References}
\begin{enumerate}
\item \label{pct} R.F.~Streater and A.S.~Wightman, {\em PCT, Spin and Statistics, and All That}, Reading, MA: W.A.~Benjamin, 1964.
\item \label{bogo} N.N.~Bogolubov, A.A.~Logunov, and I.T.~Todorov, {\em Introduction to Axiomatic Quantum Field Theory}, trans.~by Stephen Fulling and Ludmilla Popova, Reading, MA: W.A.~Benjamin, 1975.
\item \label{wightman-hilbert} A.S.~Wightman,``Hilbert's Sixth Problem: Mathematical Treatment of the Axioms of Physics'', {\em Mathematical Development Arising from Hilbert Problems}, ed.~by F.~E.~Browder, {\em Symposia in Pure Mathematics 28}, Providence, RI: Amer.~Math.~Soc., 1976, p.~147.
\item \label{wight} A.S.~Wightman, ``Quantum Field Theory in Terms of Vacuum Expectation Values'', {\em Phys.~Rev.}, vol.~101, 1956, p.~860.
\item \label{borchers} H.J.~Borchers, ``On the structure of the algebra of field operators'', {\em Nuovo Cimento}, Vol.~24, 1962, p.~214.
\item \label{gel2} I.M.~Gelfand, and G.E.~Shilov, {\em Generalized Functions, Vol.~2}, trans.~M.D.~Friedman, A.~Feinstein, and C.P.~Peltzer, New York, NY: Academic Press, 1968.
\item \label{baum} H.~Baumg\"{a}rtel, and M.~Wollenberg, ``A Class of Nontrivial Weakly Local Massive Wightman Fields with Interpolating Properties'', {\em Commun.~Math.~Phys.} Vol.~94, 1984, p.~331.
\item \label{lechner1} D.~Buchholz, G.~Lechner and S.J.~Summers, ``Warped Convolutions, Rieffel Deformations and the Construction of Quantum Field Theories'', {\em Commun.~Math.~Phys.} Vol.~304, 2011, p.~95.
\item \label{lechner2} G.~Lechner, ``Deformations of quantum field theories and integrable models'', {\em Commun. Math. Phys.} Vol.~312, 2012, p.~265.
\item \label{gej05} G.E.~Johnson, ``Algebras without Involution and Quantum Field Theories'', March 2012, arXiv:math-ph/1203.2705.
\item \label{johnson} G.E.~Johnson, ``Measurement and self-adjoint operators'', May 2014, arXiv:quant-ph/\-1405.\-7224.
\item \label{vonN} J.~von Neumann, {\em Mathematical Foundations of Quantum Mechanics}, Princeton, NJ: Princeton University Press, 1955.
\item \label{yngvason-lqp} J.~Yngvason, ``Localization and Entanglement in Relativistic Quantum Physics'', Jan.~2014, arXiv:quant-ph/1401.2652.
\item \label{wigner} T.D.~Newton and E.P.~Wigner, ``Localized States for Elementary Systems'', {\em Rev.~Modern Phys.}, Vol.~21, 1949, p.~400.
\item \label{agw} S.~Albeverio, H.~Gottschalk, and J.-L.~Wu, ``Convoluted generalized white noise, Schwinger functions and their analytic continuation to Wightman functions'', {\em Rev. Math. Phys.}, Vol.~8, 1996, pg.~763-817.
\item \label{iqf} G.E.~Johnson, ``Interacting quantum fields'', {\em Rev.~Math.~Phys.}, Vol.~11, 1999, p.~881 and errata, {\em Rev.~Math.~Phys.}, Vol.~12, 2000, p.~687.\footnote{The errata for [\ref{iqf}] is that the constructed Euclidean region semi-norm is insufficient to result in Hilbert space field operators. With the revised axioms, the $n$-point generalized functions are not necessarily the boundary values of analytic functions associated with a single random process.}
\item \label{ieee} G.E.~Johnson, ``Constructions of Particular Random Processes'', {\em Proceedings of the IEEE}, Vol. 82, no. 2, 1994, p.~270.
\item \label{feymns} G.E.~Johnson, ``Fields and Quantum Mechanics'', Dec.~2013, arXiv:math-ph/\-1312.\-2608.
\item \label{weinberg} S.~Weinberg, {\em The Quantum Theory of Fields, Volume I, Foundations}, New York, NY: Cambridge University Press, 1995.
\item \label{mp01} G.E.~Johnson, ``Massless Particles in QFT from Algebras without Involution'', May 2012, arXiv:math-ph/1205.4323.
\item \label{dieud} J.~Dieudonn\'{e}, {\em Treatise on Analysis, Volume II}, trans.~by I.G.~MacDonald, Vol.~10-II in Pure and Applied Mathematics, A Series of Monographs and Textbooks, New York, NY: Academic Press, 1970.
\item \label{haagasym} R.~Haag, ``Quantum Field Theories with Composite Particles and Asymptotic Conditions'', {\em Phys.~Rev.}, Vol.~112, 1958, p.~669.
\item \label{ahh} H.~Araki, K.~Hepp, and D.~Ruelle, ``On the Asymptotic Behavior of Wightman Functions in Space-Like Directions'', {\em Helv.~Phys.~Acta.}, Vol.~35, 1962, p.~164.
\item \label{horn} R.A.~Horn, and C.R.~Johnson, {\em Matrix Analysis}, Cambridge: Cambridge University Press, 1985.
\item \label{ruelle-link} D. Ruelle, {\em Statistical Mechanics, Rigorous Results}, Reading, MA: W.A.~Benjamin, 1969.
\item \label{combin} M.~Hardy, ``Combinatorics of Partial Derivatives'', {\em The Electronic Journal of Combinatorics}, Vol.~13, \#R1, 2006.
\item \label{nbs} {\em Handbook of Mathematical Functions}, ed.~M.~Abramowitz, and I.A.~Stegun, National Bureau of Standards, Applied Mathematics Series - 55, 1970.
\item \label{gel1} I.M.~Gelfand, and G.E.~Shilov, {\em Generalized Functions, Vol.~1}, trans.~by E.~Saletan, New York, NY: Academic Press, 1964.
\item \label{titchmarsh} E.C.~Titchmarsh, {\em Introduction to the Theory of Fourier Integrals}, Oxford: The Oxford University Press, 1948.
\item \label{reeh} H.~Reeh and S.~Schlieder, ``Bemerkungen zur Unit\"{a}r\"{a}quivalenz von Lorentz\-invarianten Feldern'', {\em Nuovo Cimento}, Vol.~22, 1961, p.~1051.
\item \label{segal} I.E.~Segal and R.W.~Goodman, ``Anti-locality of certain Lorentz-invariant operators'', {\em Journal of Mathematics and Mechanics}, Vol.~14, 1965, p.~629.
\item \label{masuda} K.~Masuda, ``Anti-Locality of the One-half Power of Elliptic Differential Operators'', {\em Publ. RIMS, Kyoto Univ.}, Vol.~8, 1972, p.~207.
\item \label{heger} G.C.~Hegerfeldt, ``Prime Field Decompositions and Infinitely Divisible States on Borchers' Tensor Algebra'', {\em Commun.~Math.~Phys.}, Vol.~45, 1975, p.~137.
\item \label{feder} P.G.~Federbush and K.A.~Johnson, ``The Uniqueness of the Two-Point Function'', {\em Phys.~Rev.}, Vol.~120, 1960, p.~1926.
\item \label{greenberg} O.W.~Greenberg, ``Heisenberg Fields which vanish on Domains of Momentum Space'', {\em Journal of Math.~Phys.}, Vol.~3, 1962, pp.~859-866.
\item \label{ruelle} D.~Ruelle, ``On the Asymptotic Condition in Quantum Field Theory'', {\em Helv. Phys. Acta.}, Vol.~35, 1962, p.~147.
\item \label{gel4} I.M.~Gelfand, and N.Ya.~Vilenkin, {\em Generalized Functions, Vol.~4}, trans.~A.~Feinstein, New York, NY: Academic Press, 1964.
%
\end{enumerate}
\end{document}